\documentclass[
 reprint,
 amsmath,amssymb,
 aps,
 pre,
]{revtex4-2}

\usepackage{graphicx}
\usepackage{dcolumn}
\usepackage{bm}

\begin{document}

\preprint{APS/123-QED}

\title{Statistical Analysis of Speckle Fields}

\author{Ian D. Min-Roberts}%

\author{Wojciech Rozmus}%
\affiliation{Department of Physics, University of Alberta, Edmonton, Alberta, Canada}

\author{Pierre A. Michel}%
\affiliation{Lawrence Livermore National Laboratory, Livermore, California, USA}

\date{\today}

\begin{abstract}
Speckles, or laser hot spots, are the intensity maxima of optically smoothed laser beams that seed instabilities, making a quantitative statistical description of speckles of interest. Earlier statistical theories estimated the number of speckles above an intensity level set for optically smoothed beams produced by random phase plates, using an ansatz that relates intensity maxima to the maxima of the real and imaginary components of the underlying complex Gaussian electric field. Here, we count speckles directly from the laser intensity field, treating it as a $\chi_2^2$ random field and imposing the local maximum conditions without a single-component ansatz. We evaluate the theory for square, circular, annular, and Gaussian aperture spectra, and include induced spatial incoherence as a temporal smoothing mechanism. Monte Carlo simulations confirm the theory and show improved accuracy relative to the earlier ansatz-based formulation. Applications include a simple SBS reflectivity model using the resulting speckle statistics, and a comparison between speckle-driven and thermal noise density-fluctuation spectra.
\end{abstract}


\maketitle


\section{\label{sec:Introduction}Introduction}
High-energy laser experiments require intense laser beams with controlled spatial and temporal structure at the target. Optical smoothing techniques can be implemented to manipulate the spatial and temporal coherence properties of laser beams to reduce the effects of optical aberrations and large-scale intensity non-uniformities. Implementations include random phase plates \cite{kato1984random} (RPP), induced spatial incoherence \cite{lehmberg1987theory} (ISI), and smoothing by spectral dispersion \cite{skupsky1989improved} (SSD). Despite reducing large-scale intensity non-uniformities, these techniques produce a complex interference pattern in the focal region consisting of localized regions of enhanced intensity commonly referred to as speckles \cite{michel2023introduction}. The intensity field may then be viewed as a random collection of speckles with characteristic spatial and temporal scales determined by the underlying beam smoothing technique.

A quantity of particular interest is the expected number of speckles whose intensities exceed a prescribed threshold, since such speckles often dominate nonlinear laser-plasma interactions. Statistical theories for such speckles were developed by Rose \& DuBois~\cite{rose1993statistical} and later corrected by Garnier~\cite{garnier1999statistics}. In both cases, the counting problem is formulated using a single component ansatz, where the expected number of speckles above a given threshold is estimated from a single Gaussian component of the underlying complex electric field. Since speckles are fundamentally defined through the intensity field, it is natural to formulate the counting problem directly in terms of the intensity. The present work builds on these earlier developments by revisiting the counting problem from this perspective. Specifically, the laser intensity field is treated as a $\chi_2^2$ random field, allowing local maxima theory for $\chi_2^2$ random fields to be applied directly. This yields an asymptotic formula for the expected number of speckles whose intensities exceed a given level set.

We begin by introducing the paraxial laser model, aperture boundary conditions, and the statistical descriptions of the vector potential and intensity fields. These results are then used to derive a new asymptotic counting theory for intense speckles across several beam models, including square, circular, and annular RPP beams, as well as Gaussian and ISI beams. We next establish that the characteristic transverse and longitudinal speckle dimensions are determined by the second-order spectral moments of the intensity spectrum. For ease of reference, key analytical results are collected into equations and summary tables. The resulting counting formula is then assessed against the single-component ansatz of Rose and Garnier. Finally, we apply the theory to average SBS reflectivity and speckle-induced density fluctuations, and show how annular apertures provide additional control over speckle size.

\section{\label{sec:Laser_Beam_Model}Laser Beam Model}
The wave equation for a propagating laser field can be described by the vector potential $\mathbf{A}\colon\mathbb{R}^3\times\mathbb{R}_+\to\mathbb{R}^3,$ where the domain coordinates are $(\mathbf{x}_{\perp},z,t)\in\mathbb{R}^3\times\mathbb{R}_+$, with $\mathbf{x}_{\perp}\in\mathbb{R}^2$ denoting the transverse laser directions, $z\in\mathbb{R}$ denoting the propagation axis, and $t\in\mathbb{R}_+$ denoting time. For constant plasma frequency $\omega_{pe}$, it can be written as
\begin{equation}
\nabla^2\mathbf{A} - \frac{1}{c^2}\frac{\partial^2\mathbf{A}}{\partial t^2} = \frac{\omega_{pe}^2}{c^2} \mathbf{A}.
\label{eq:wave_equation}
\end{equation}
We then represent the vector potential by its slowly varying complex envelope $\mathcal{A}\colon\mathbb{R}^3\times\mathbb{R}_+\to\mathbb{C}$
\begin{equation}
\mathbf{A}(\mathbf{x}_{\perp},z,t) = \Re\left\lbrace \mathcal{A}(\mathbf{x}_{\perp},z,t) e^{i\left(k_0z-\omega_0t\right)}\hat{\mathbf{e}}_p\right\rbrace,
\label{eq:envelope}
\end{equation}
where $\hat{\mathbf{e}}_p$ is a constant unit polarization vector, $\omega_0$ is the carrier frequency, and $k_0$ is the carrier wavenumber associated with the background plasma frequency $\omega_{pe}$, satisfying $c^2k_0^2 = \omega_0^2-\omega_{pe}^2$. Substituting Eq.~\eqref{eq:envelope} into Eq.~\eqref{eq:wave_equation}, applying the slowly-varying-envelope approximation: $\lvert \partial_z^2\mathcal{A} \rvert \ll \lvert k_0\partial_z\mathcal{A} \rvert$ and $\lvert \partial_t^2\mathcal{A} \rvert \ll \lvert \omega_0\partial_t\mathcal{A} \rvert$, and shifting to the retarded time coordinate $t\mapsto t-z/v_g$ for $v_g=c^2k_0/\omega_0$, yields the paraxial equation
\begin{equation}
\begin{cases}
2ik_0\partial_z\mathcal{A} + \nabla_{\perp}^2\mathcal{A} = 0,
& (\mathbf{x}_{\perp},z,t)\in\mathbb{R}^4,\\
\mathcal{A}(\mathbf{x}_{\perp},0,t)=\mathcal{A}_0(\mathbf{x}_{\perp},t),
& (\mathbf{x}_{\perp},t)\in\mathbb{R}^3,
\end{cases}
\label{eq:paraxial_pde}
\end{equation}
where $\mathcal{A}_0(\mathbf{x}_{\perp},t)$ is the entrance plane boundary envelope, $\nabla_{\perp}^2$ is the Laplacian with respect to $\mathbf{x}_{\perp}$ coordinates, and now $t\in\mathbb{R}$ is the retarded time coordinate.

\subsection{\label{sec:Boundary_Condition}Boundary Condition}
The boundary condition $\mathcal{A}_0$ describes the laser field in the near field after spatial and temporal smoothing. Generalized beamlet representations of the following form have been widely employed in studies of laser-plasma instabilities, including stimulated Brillouin scattering, filamentation, and cross-beam energy transfer \cite{ruyer2023influence,ruyer2023backward,ruyer2025statistical,oudin2025theory}. Let $\mathfrak{a}$ denote a square-summable unit-normalized spectral weighting function associated with the aperture and pulse structure, then
\begin{equation}
\mathcal{A}_0(\mathbf{x}_\perp,t) = \sigma_{\mathcal{A}\mathcal{A}}\sum_{\mathbf{k}_{\perp},\omega} \mathfrak{a}(\mathbf{k}_\perp,\omega)e^{i\varphi(\mathbf{k}_\perp,\omega)} e^{i(\mathbf{k}_{\perp}\cdot\mathbf{x}_{\perp}-\omega t)}.
\label{boundary_condition}
\end{equation}
where $\sigma_{\mathcal{A}\mathcal{A}}$ is the vector potential amplitude, and $e^{i\varphi(\mathbf{k}_\perp,\omega)}$ represents the phase modulation introduced by the beam-smoothing optics. 

\subsubsection{Phase Modulator}
For a RPP beam, $\varphi(\mathbf{k}_\perp,\omega) = \varphi_{\mathrm{RPP}}(\mathbf{k}_{\perp})$, where independent binary phases $\{0,\pi\}$ are assigned to the transverse modes with equal probability, or equivalently $e^{i\varphi_{\mathrm{RPP}}(\mathbf{k}_{\perp})}\in\{-1,+1\}$ with equal probability. For an ISI beam, $\varphi(\mathbf{k}_\perp,\omega) = \varphi_{\mathrm{ISI}}(\mathbf{k}_{\perp},\omega)$, where $e^{i\varphi_{\mathrm{ISI}}(\mathbf{k}_{\perp},\omega)}\in\mathbb{S}^1$ is modelled as uniformly distributed on the unit circle in the complex plane. When both smoothing mechanisms are present, we take
\[
\varphi(\mathbf{k}_\perp,\omega) = \varphi_{\mathrm{RPP}}(\mathbf{k}_{\perp}) + \varphi_{\mathrm{ISI}}(\mathbf{k}_{\perp},\omega).
\]
Since the RPP and ISI phase modulations are taken to be statistically independent,
\begin{equation}
\left\langle e^{i\varphi(\mathbf{k}_\perp,\omega)} \right\rangle = \left\langle e^{i\varphi_{\mathrm{RPP}}(\mathbf{k}_{\perp})} \right\rangle \left\langle e^{i\varphi_{\mathrm{ISI}}(\mathbf{k}_{\perp},\omega)} \right\rangle = 0.
\label{eq:zero_mean_phasors}
\end{equation}
We also assume that the phasors are uncorrelated between distinct discrete modes, so that
\begin{equation}
\left\langle e^{i\varphi(\mathbf{k}_\perp,\omega)} \overline{e^{i\varphi(\mathbf{k}_\perp',\omega')}} \right\rangle = \delta_{\mathbf{k}_\perp,\mathbf{k}'_\perp} \delta_{\omega,\omega'},
\label{eq:phasor_covariance}
\end{equation}
where $\delta_{\mathbf{k}_\perp,\mathbf{k}'_\perp}$ and $\delta_{\omega,\omega'}$ are Kronecker deltas.

\begin{figure}
\includegraphics[width=\linewidth]{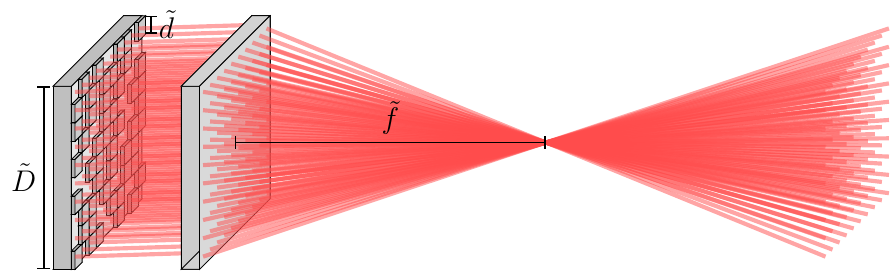}
\caption{A laser beam propagating through a square RPP and lens. Here $\tilde{f}$ is the focal length, $\tilde{D}$ is the aperture diameter, and $\tilde{d}$ is an RPP element width. Here the tilde accent denotes measurable physical parameters of the optical system.}
\label{fig:model} 
\end{figure}

\subsubsection{Spectral Weighting Function}
The spectral weighting function $\mathfrak{a}(\mathbf{k}_\perp,\omega)$ describes the distribution of the optical field variance among the spatial and temporal modes supported by the smoothing technique. For a RPP beam, the spectral weighting function is independent of frequency and may be written as $\mathfrak{a}(\mathbf{k}_\perp,\omega) = \mathfrak{a}_{\perp}(\mathbf{k}_\perp)$. Let $\tilde{D}$ denote the aperture diameter and $\tilde{d}$ the phase element width, as shown in Fig.~\ref{fig:model}. The aperture diameter determines the transverse wave-number scale of the beam, while the phase element width determines the spacing between neighbouring transverse modes. These are
\begin{equation}
k_c = \frac{k_0\tilde{D}}{\sqrt{\tilde{D}^{\,2}+4\tilde{f}^{2}}} \approx \frac{k_0\tilde{D}}{2\tilde{f}}, \qquad \tilde f \gg \tilde D,
\label{wavenumber_cutoff}
\end{equation}
and
\begin{equation}
\Delta_{k_\perp} = \frac{2k_0\tilde{d}}{\sqrt{\tilde{d}^{\,2}+4\tilde{f}^{\,2}}} \approx \frac{k_0\tilde{d}}{\tilde{f}}, \qquad \tilde f \gg \tilde d ,
\label{transverse_mode_spacing}
\end{equation}
where $\tilde{f}$ is the focal length of the optic. In high-energy laser systems, the aperture spectrum is often well approximated by a top-hat profile, corresponding to a nearly uniform distribution of optical power across the aperture. We consider square and circular top-hat models, as well as an annular top-hat model. The annular model is intended to provide a useful idealized geometry for optical configurations where many laser beams are arranged into a cone at a fixed polar angle around a symmetry axis, like in indirect-drive ICF lasers such as the National Ignition Facility (NIF) or Laser Megajoule (LMJ). We also consider a Gaussian aperture. Unlike the top-hat models, the Gaussian aperture does not possess compact support, but provides a useful analytical approximation whose width may be chosen to reproduce characteristic speckle scales associated with physical optical systems.

\begin{figure*}
\includegraphics[width=0.23\linewidth]{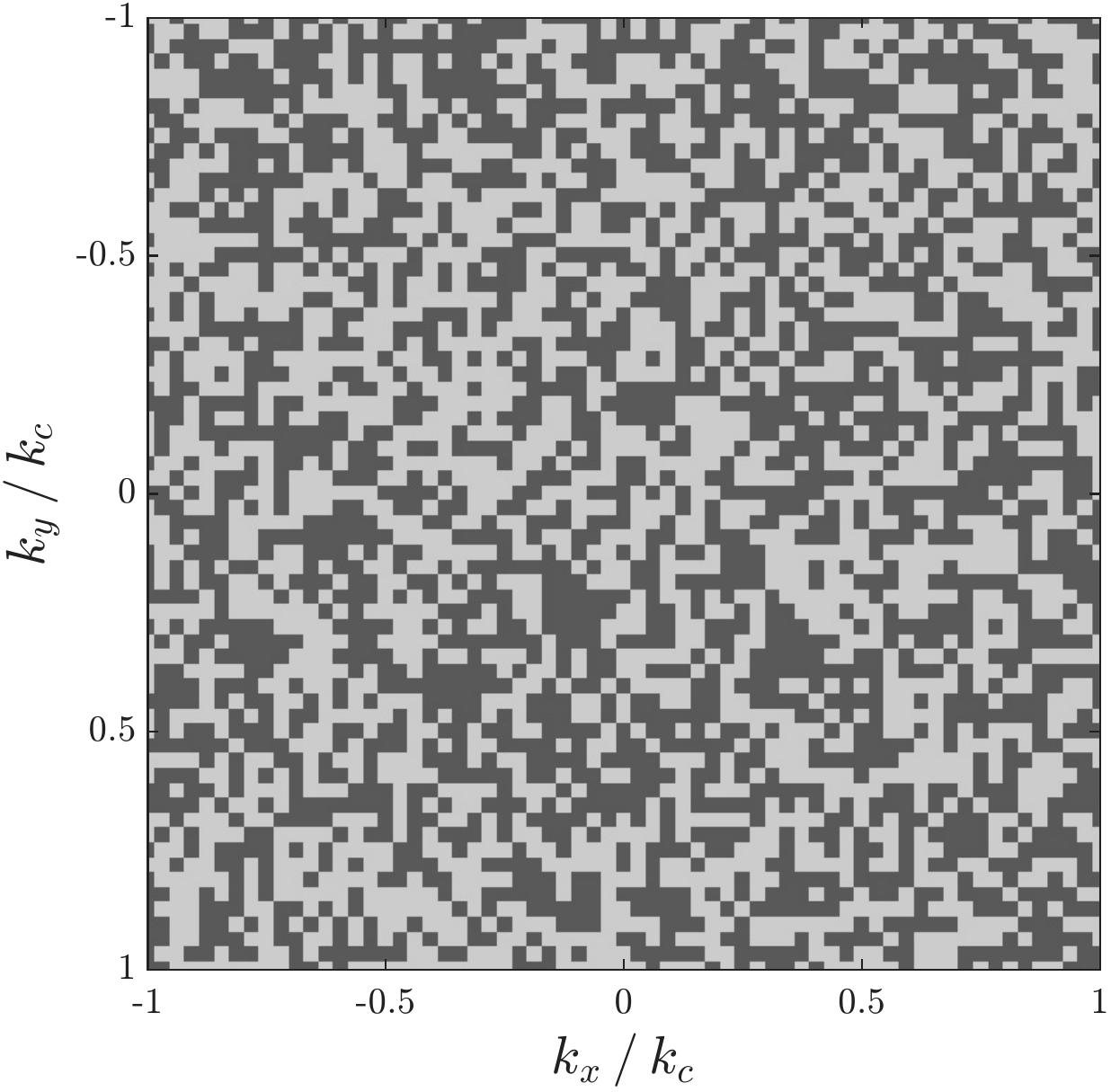}
\hfill
\includegraphics[width=0.23\linewidth]{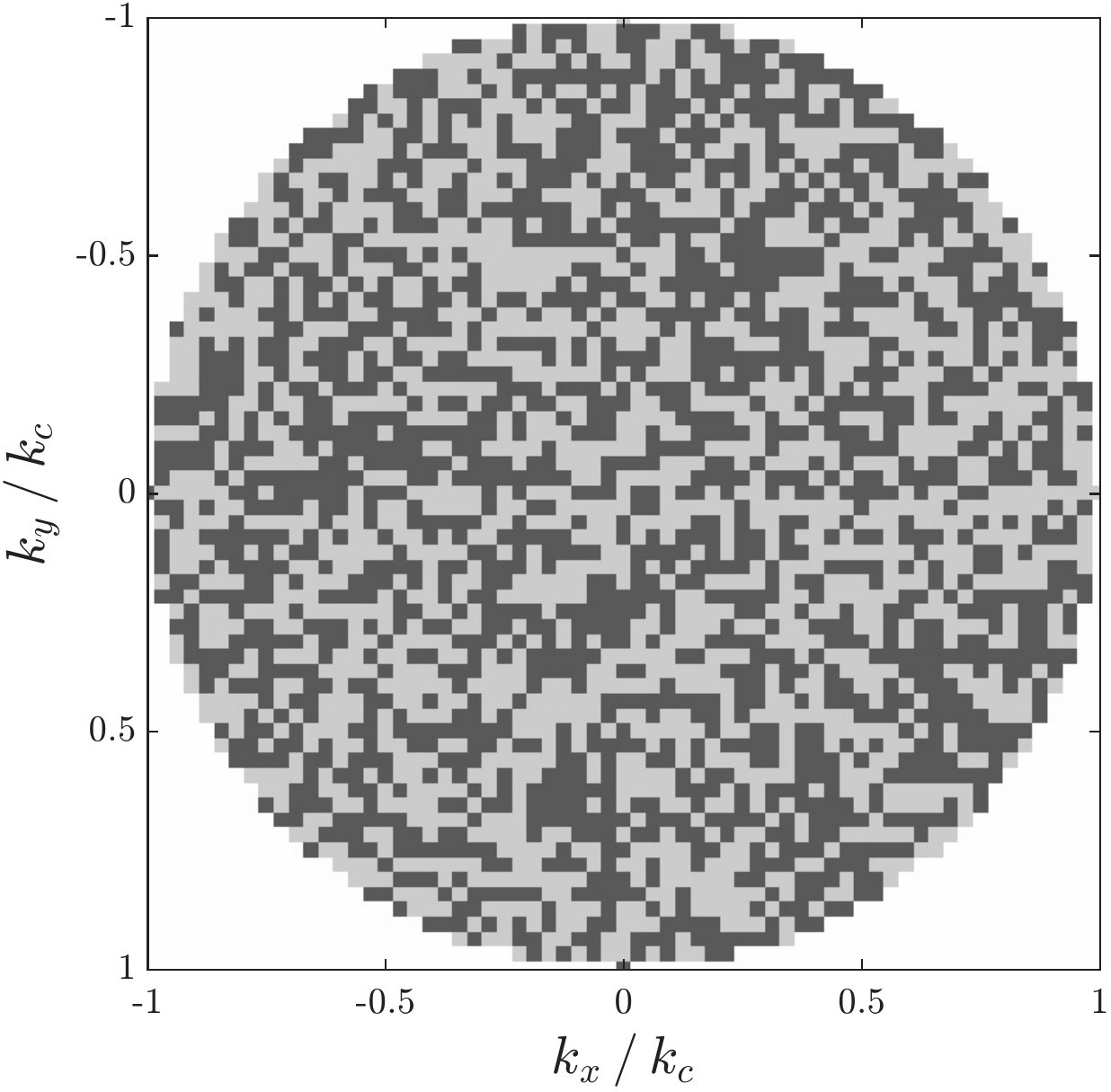}
\hfill
\includegraphics[width=0.23\linewidth]{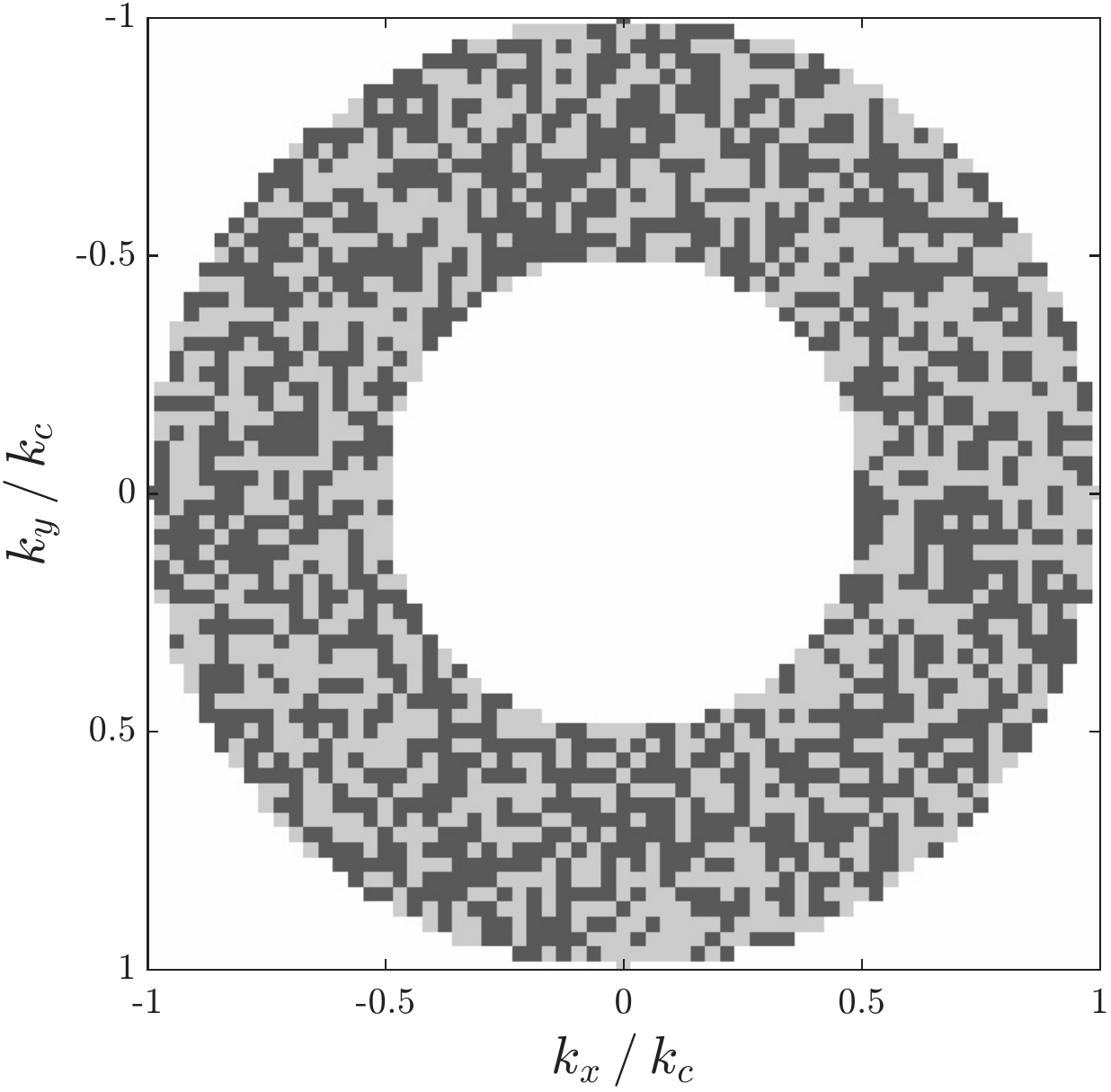}
\hfill
\includegraphics[width=0.23\linewidth]{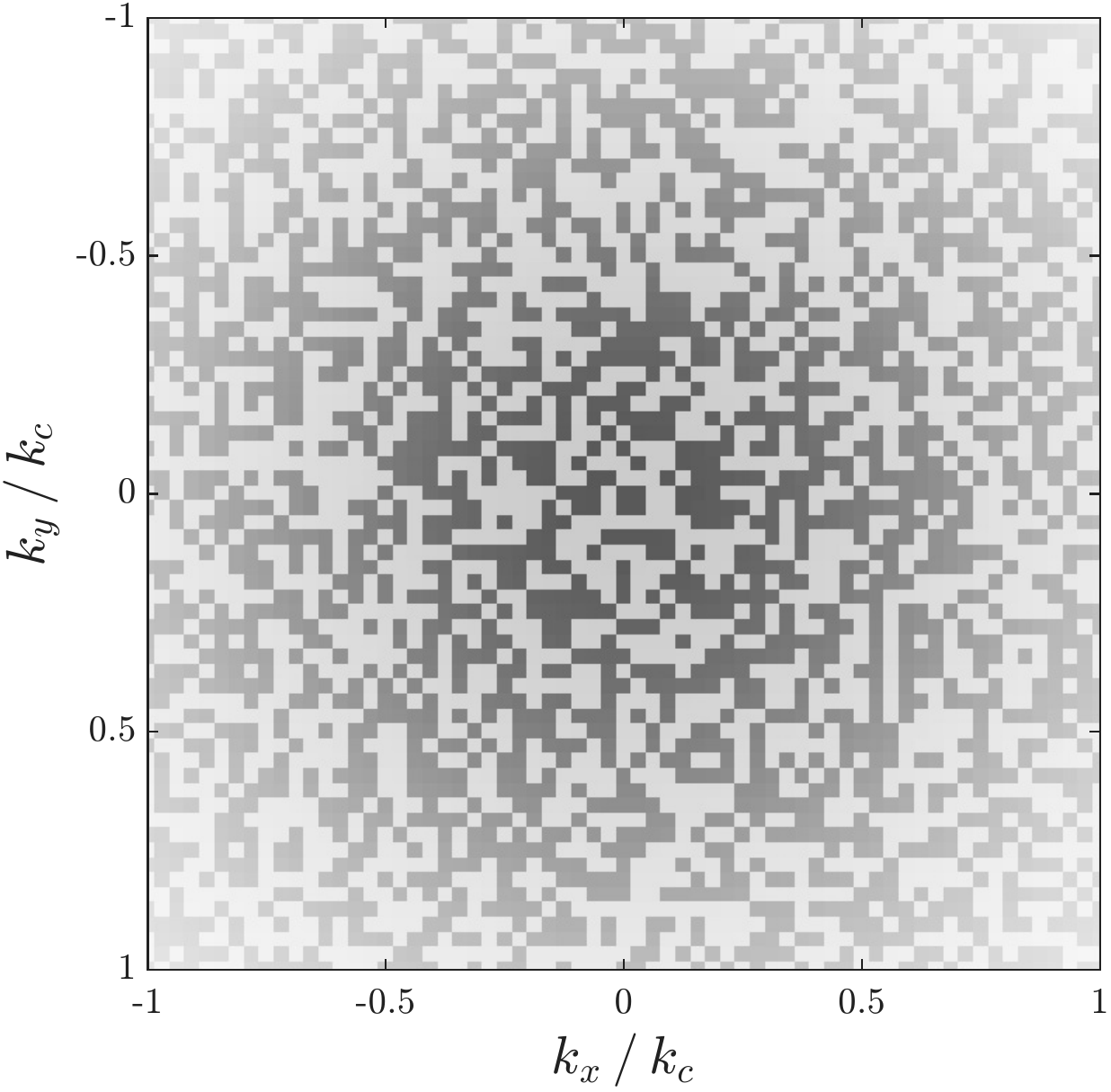}
\caption{RPP design for the four aperture models, square aperture, circular aperture, annular aperture, and Gaussian aperture, respectively. Note the Gaussian aperture extends past $k_c$ since it has no compact support.}
\label{fig:aperture_models}
\end{figure*}

Let $\mathbf{1}_X$ denote the indicator function on a set $X$. With the discrete transverse modes separated by $\Delta_{k_\perp}$, the unit-normalized spatial mode weights are
\begin{align}
\text{Square:} &\quad \left\lvert\mathfrak{a}_{\perp\square}\right\rvert^2 = \frac{\mathbf{1}_{[-k_c,k_c]^2}}{(2k_c)^2}\, \Delta_{k_\perp}^2, \label{boundary_square}\\
\text{Circular:} &\quad \left\lvert\mathfrak{a}_{\perp\circ}\right\rvert^2 = \frac{\mathbf{1}_{\{k_{\perp}\leq k_c\}}}{\pi k_c^2}\, \Delta_{k_\perp}^2, \label{boundary_circular}\\
\text{Annular:} &\quad \left\lvert\mathfrak{a}_{\perp\circledcirc}\right\rvert^2 = \frac{\mathbf{1}_{\{k_I\leq k_{\perp} \leq k_O\}}}{\pi(k_O^2-k_I^2)}\, \Delta_{k_\perp}^2, \label{boundary_annular}\\
\text{Gaussian:} &\quad \left\lvert\mathfrak{a}_{\perp g}\right\rvert^2 = \frac{L_\perp^2}{2\pi}\exp\left(-\frac{L_\perp^2 k_{\perp}^2}{2}\right)\Delta_{k_\perp}^2, \label{boundary_gaussian}
\end{align}
where $k_{\perp}=|\mathbf{k}_{\perp}|$. For ISI, the spectral weighting function includes both spatial and temporal smoothing. These contributions are separable, so that
\begin{equation}
\mathfrak{a}(\mathbf{k}_\perp,\omega) = \mathfrak{a}_{\perp}(\mathbf{k}_\perp)\mathfrak{a}_{t}(\omega).
\end{equation}
The temporal weighting is taken to be Gaussian with coherence time $\ell_t$. With temporal modes separated by $\Delta_\omega$, the unit-normalized temporal mode weights are
\begin{equation}
\left|\mathfrak{a}_{t}(\omega)\right|^2 = \frac{\ell_t}{\sqrt{2\pi}} \exp\left(-\frac{\ell_t^2\omega^2}{2}\right) \Delta_\omega.
\label{boundary_temporal}
\end{equation}

\subsection{\label{sec:Paraxial_Solution}Paraxial Solution}
To determine the propagation of the boundary field, consider a single spectral mode $(\mathbf{k}_\perp,\omega)$ in our boundary condition, Eq.~\eqref{boundary_condition}, which we write as
\[
\mathcal A_{\mathbf{k}_\perp,\omega}(\mathbf{x}_\perp,z,t) = \psi_{\mathbf{k}_\perp}(z) e^{i(\mathbf{k}_\perp\cdot\mathbf{x}_\perp-\omega t)},
\]
such that $\psi_{\mathbf{k}_\perp}(0)=1$. Substitution into the paraxial Eq.~\eqref{eq:paraxial_pde} gives us that
\begin{equation}
\psi_{\mathbf{k}_\perp}(z) = \exp\left(-i\frac{k_\perp^2 z}{2k_0}\right).
\label{eq:propagator}
\end{equation}
where $k_{\perp}=|\mathbf{k}_{\perp}|$. Then by linearity, the full solution to our paraxial Eq.~\eqref{eq:paraxial_pde} is
\begin{align}
\mathcal{A}(\mathbf{x}_\perp,z,t) =&\, \sigma_{\mathcal{A}\mathcal{A}} \sum_{\mathbf{k}_{\perp},\omega} \mathfrak{a}(\mathbf{k}_\perp,\omega) e^{i\varphi(\mathbf{k}_\perp,\omega)} \nonumber\\
&\times \exp\!\left(i(\mathbf{k}_{\perp}\cdot\mathbf{x}_{\perp}-\omega t)-i\frac{k_\perp^2 z}{2k_0}\right).
\label{paraxial_solution}
\end{align}

\subsubsection{Covariance}
In Eq.~\eqref{paraxial_solution}, the vector potential is random because of the phase modulation $e^{i\varphi(\mathbf{k}_\perp,\omega)}$. Therefore, rather than focusing on a single realization of the field, we characterize its physical structure statistically through the covariance function, defined by
\begin{equation}
C_{\mathcal{A}\mathcal{A}} = \left\langle \mathcal{A}(\mathbf{x}_\perp,z,t) \overline{\mathcal{A}(\mathbf{x}_\perp',z',t')}\right\rangle.
\end{equation}
Substituting in Eq.~\eqref{paraxial_solution}, we get that the covariance can be written as
\begin{align}
C_{\mathcal{A}\mathcal{A}} =&\, \sigma_{\mathcal{A}\mathcal{A}}^{2} \sum_{(\mathbf{k}_{\perp},\omega)} \sum_{(\mathbf{k}'_{\perp},\omega')} \mathfrak{a}(\mathbf{k}_{\perp},\omega) \overline{\mathfrak{a}(\mathbf{k}'_{\perp},\omega')} \nonumber\\
&\times \left\langle e^{i\varphi(\mathbf{k}_\perp,\omega)} \overline{e^{i\varphi(\mathbf{k}_\perp',\omega')}} \right\rangle \psi(\mathbf{k}_{\perp},z) \overline{\psi(\mathbf{k}'_{\perp},z')}
\nonumber\\
&\times e^{i(\mathbf{k}_{\perp}\cdot\mathbf{x}_{\perp}-\omega t)} e^{-i(\mathbf{k}'_{\perp}\cdot\mathbf{x}'_{\perp}-\omega' t')}.
\label{eq:covariance_expanded}
\end{align}
Then using the Kronecker delta relations for our phase modulator Eq.~\eqref{eq:phasor_covariance}, substituting in the propagator Eq.~\eqref{eq:propagator}, and noticing that the covariance only depends on coordinate differences so that we can shift our coordinate system $(\mathbf{x}_{\perp}-\mathbf{x}_{\perp}^\prime, z-z^\prime, t-t^\prime ) \mapsto (\mathbf{x}_\perp,z,t)$, we arrive at the result
\begin{equation}
C_{\mathcal{A}\mathcal{A}}(\mathbf{x}_\perp,z,t) = \sigma_{\mathcal{A}\mathcal{A}}^{2} \sum_{\mathbf{k}_{\perp},\omega} \left\lvert \mathfrak{a}(\mathbf{k}_{\perp},\omega) \right\rvert^{2} e^{i(\mathbf{k}_{\perp}\cdot\mathbf{x}_{\perp}-\omega t) -i\frac{k_{\perp}^{2}z}{2k_0}}.
\label{eq:covariance_discrete}
\end{equation}

\subsubsection{Spectral Density}
The covariance Eq.~\eqref{eq:covariance_discrete} is expressed as a weighted sum over discrete spectral modes. Since the mode weights contain the spectral cell dimensions $\Delta_{k_\perp}^2$ and $\Delta_{\omega}$, it is convenient to separate these factors by introducing the boundary spectral density $S_{\mathcal{A}_0\mathcal{A}_0}$ through
\begin{equation}
\sigma_{\mathcal{A}\mathcal{A}}^{2}\left\lvert \mathfrak{a}(\mathbf{k}_{\perp},\omega) \right\rvert^{2} = S_{\mathcal{A}_0\mathcal{A}_0}(\mathbf{k}_{\perp},\omega) \frac{\Delta_{k_\perp}^2\Delta_{\omega}}{(2\pi)^3}.
\label{eq:spectral_density_A}
\end{equation}
This separates the variance density from the finite spectral cell size. When many independently phased modes contribute to the field, the discrete mode weights may be viewed as samples of a continuous boundary spectrum. For a RPP beam, this corresponds to $\tilde{d}\ll \tilde{D}$, or equivalently $2k_c/\Delta_{k_\perp}=\tilde{D}/\tilde{d}\gg1$. The cell volume then becomes the Fourier integration element, $\Delta_{k_\perp}^2\Delta_\omega\rightarrow d^2\mathbf{k}_\perp d\omega$, giving
\begin{align*}
C_{\mathcal{A}\mathcal{A}}(\mathbf{x}_\perp,z,t) =& \frac{1}{(2\pi)^3}\int_{\mathbb{R}^{3}} S_{\mathcal{A}_0\mathcal{A}_0}(\mathbf{k}_{\perp},\omega)\\ 
& \times e^{i(\mathbf{k}_{\perp}\cdot\mathbf{x}_{\perp}-\omega t)-i\frac{k_{\perp}^{2}z}{2k_0}}\,d^2\mathbf{k}_{\perp}\, d\omega,
\end{align*}
which can be conveniently expressed as
\begin{equation}
C_{\mathcal{A}\mathcal{A}}(\mathbf{x}_\perp,z,t) = \mathcal{F}^{-1}\left[S_{\mathcal{A}_0\mathcal{A}_0}(\mathbf{k}_{\perp},\omega) e^{-i\frac{k_{\perp}^{2}z}{2k_0}}\right],
\label{eq:covariance_spectral}
\end{equation}
where $\mathcal{F}^{-1}\colon L^2(\mathbb{R}^3)\to L^2(\mathbb{R}^3)$ denotes the inverse Fourier transform mapping $(\mathbf{k}_{\perp},\omega)\mapsto(\mathbf{x}_{\perp},t)$ under the non-unitary convention.\\

By construction, $C_{\mathcal{A}\mathcal{A}}$ and $S_{\mathcal{A}_0\mathcal{A}_0}$ are normalized so that $C_{\mathcal{A}\mathcal{A}}(0,0,0)=\sigma_{\mathcal{A}\mathcal{A}}^2$. In the following, we will work primarily with the corresponding unit-normalized covariance and spectral density. For notational convenience, we define
\begin{equation}
C = \frac{C_{\mathcal{A}\mathcal{A}}}{\sigma_{\mathcal{A}\mathcal{A}}^2}, \qquad S = \frac{S_{\mathcal{A}_0\mathcal{A}_0}}{\sigma_{\mathcal{A}\mathcal{A}}^2}.
\end{equation}

\section{\label{sec:Statistical_Analysis_of_Speckles}Statistical Analysis of Speckles}
Having established a spectral representation for the optical field, we now turn to the statistical properties of laser speckles. These properties are determined by the statistical structure of the underlying vector potential and include the field and intensity distributions, local maxima statistics, and the resulting speckle count.

\subsection{\label{sec:Gaussian_Limit_of_the_Vector_Potential}Gaussian Limit of the Vector Potential}
We first justify treating the vector potential as a complex Gaussian random field. The vector potential $\mathcal{A}$ is constructed as a sum of many independently phased spectral modes. As the number of contributing modes increases, the central limit theorem implies that $\mathcal{A}$ converges to a complex Gaussian random field. For a RPP beam, this corresponds to the physically relevant regime $\tilde{d}\ll\tilde{D}$, where many phase elements span the aperture. Furthermore, since $\langle e^{i\varphi(\mathbf{k}_\perp,\omega)}\rangle = 0$, as shown in Eq.~\eqref{eq:zero_mean_phasors}, the resulting field is zero mean. Therefore, $\mathcal{A}$ may be regarded as a mean-zero homogeneous complex Gaussian random field with covariance function $C_{\mathcal{A}\mathcal{A}}$, given by Eq.~\eqref{eq:covariance_spectral}. By the spectral representation theorem \cite[Theorem 5.4.2]{adler2007random}, the limiting Gaussian field admits an equivalent stochastic integral representation; the general construction is given in Appendix~\ref{sec:Some_Math_Definitions}, Eq.~\eqref{complex_noise}.

\subsection{\label{sec:Intensity_Statistics}Intensity Statistics}
\begin{figure}[b]
\includegraphics[width=\linewidth]{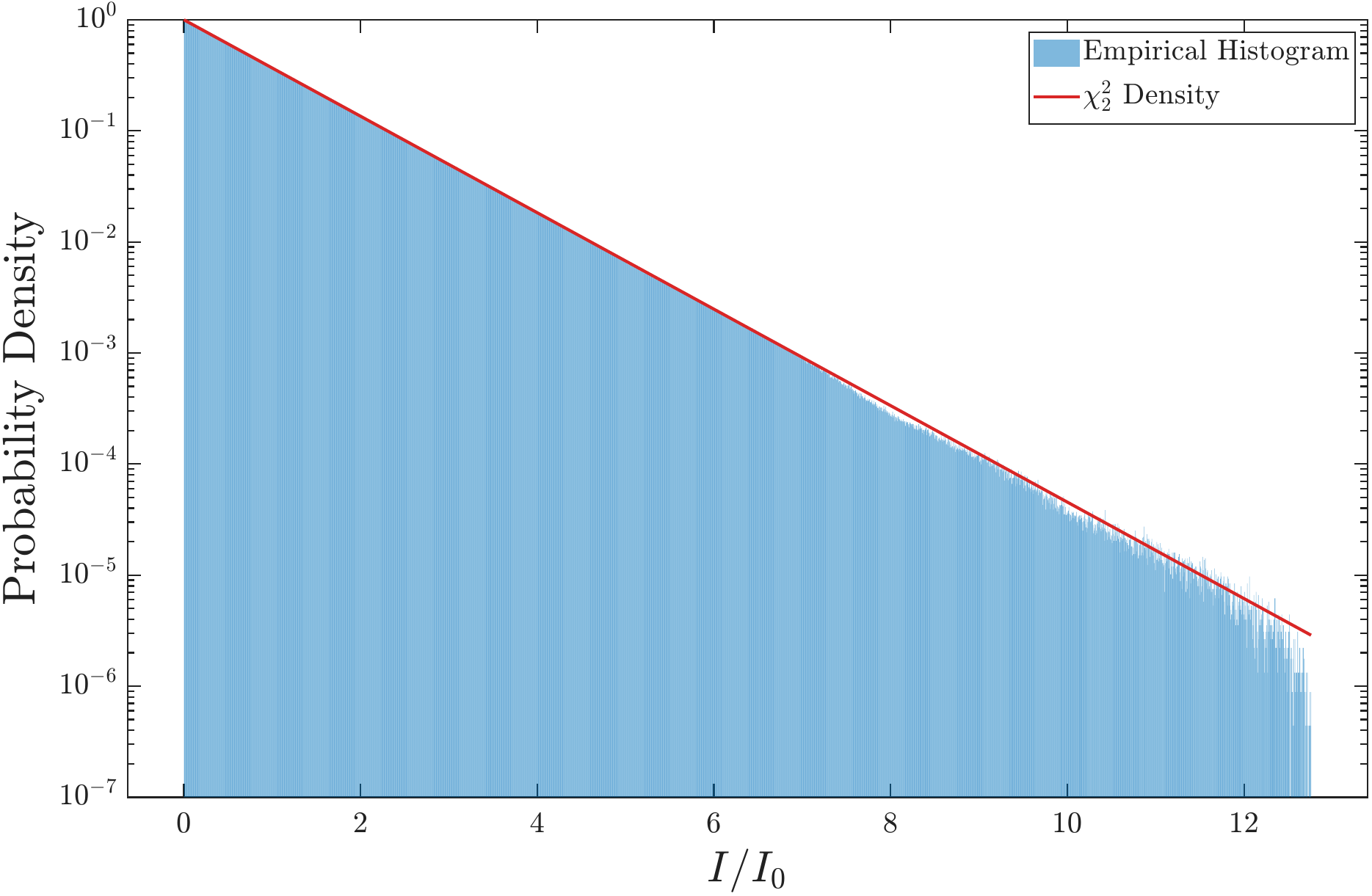}
\caption{Intensity probability density histogram simulation of one realization compared with the theoretical $\chi^2_2$ distribution given in Eq.~\eqref{intensity_pdf}, plotted on a logarithmic scale.}
\label{fig:intensity_pdf}
\end{figure}
We now define the intensity field by
\begin{equation}
I := \frac{1}{2}\varepsilon_0 c\, \omega_0^2|\mathcal{A}|^2 = \frac{1}{2}\varepsilon_0 c\, \omega_0^2\left(\mathcal{A}_R^2+\mathcal{A}_I^2\right),
\label{intensity_definition}
\end{equation}
where $\mathcal{A}=\mathcal{A}_R+i\mathcal{A}_I$. Since $\mathcal{A}$ is a proper complex Gaussian field, $\mathcal{A}_R$ and $\mathcal{A}_I$ are real Gaussian random fields. While the fields are pointwise independent, $\langle \mathcal{A}_R\, \mathcal{A}_I \rangle=0$, the propagation factor induces nonzero correlations between one field and the derivatives of the other, e.g., $\langle \nabla \mathcal{A}_R\, \mathcal{A}_I \rangle\not=0$. We require vanishing field gradient cross correlations since they enter as conditional expectations in Worsley’s Kac-Rice formula \cite[Theorem 2.1]{worsley1994local}. Thus, pointwise independence alone is not sufficient for counting local maxima above a given level set. Fortunately, the intensity field is invariant under multiplication of $\mathcal{A}$ by a unit-modulus phase factor. We therefore introduce the phase corrected field satisfying $|\mathring{\mathcal{A}}(\mathbf{x})|^2 = |\mathcal{A}(\mathbf{x})|^2$, as
\begin{equation}
\mathring{\mathcal{A}}(\mathbf{x}) = e^{-i\Lambda^{(1)}\cdot\mathbf{x}}\mathcal{A}(\mathbf{x}),
\label{phase_shift}
\end{equation}
where $\Lambda^{(1)}$ is the first order spectral moment defined in Appendix~\ref{sec:Some_Math_Definitions}, Eq.~\eqref{spectral_moment_tensor_definition}. Writing $\mathring{\mathcal{A}}=\mathring{\mathcal{A}}_R+i\mathring{\mathcal{A}}_I$, the same intensity defined in Eq.~\eqref{intensity_definition} can be rewritten as
\begin{equation}
I = \frac{1}{2}\varepsilon_0 c\, \omega_0^2\left(\mathring{\mathcal{A}}_R^2+\mathring{\mathcal{A}}_I^2\right),
\label{intensity_definition_mathring}
\end{equation}
where now $\mathring{\mathcal{A}}_R$ and $\mathring{\mathcal{A}}_I$ are pointwise independent homogeneous real Gaussian random fields with vanishing cross-correlations between one field and the derivatives of the other, giving the structure needed for the Kac-Rice framework. Thus with $I_0=\frac{1}{2}\varepsilon_0 c\, \omega_0^2\sigma_{\mathcal{A}\mathcal{A}}^2$, obtained by averaging both sides of Eq.~\eqref{intensity_definition_mathring}, we get that the intensity is distributed as chi-squared with 2 degrees of freedom. Formally we say
\begin{equation}\label{intensity_pdf}
I \sim \frac{I_0}{2}\chi_2^2, \qquad \text{with PDF} \qquad f_I(I) = \frac{1}{I_0}\,e^{-\frac{I}{I_0}}.
\end{equation}

Now that our underlying fields have been phase corrected, we must use the phase corrected second-order spectral moment~\footnote{While Rose \& DuBois~\cite{rose1993statistical} had not considered this shift due to the nature of their ansatz, Garnier~\cite{garnier1999statistics} introduced the corresponding effective-field correction for the RPP focal-volume problem. The present formulation generalizes this correction in terms of the spectral moments of a homogeneous complex Gaussian field.}. Let $\Lambda^{(k)}$ be the $k^{\text{th}}$ spectral moments defined for the underlying field $\mathcal{A}$, then the second-order spectral moment $\mathring{\Lambda}^{(2)}$ defined for the underlying field $\mathring{\mathcal{A}}$ can be written as
\begin{equation}
\mathring{\Lambda}^{(2)} = \Lambda^{(2)} - \Lambda^{(1)}\otimes\Lambda^{(1)}.
\label{eq:effective_normalized_second_order_spectral_moment}
\end{equation}
In index notation, it can be written in terms of the unit-normalized covariance as
\[
\mathring{\Lambda}_{ij}^{(2)} = - \left. \frac{\partial^2 C(\mathbf{x})}{\partial x_i\partial x_j} \right\vert_{\mathbf{x}=0} + \left. \frac{\partial C(\mathbf{x})}{\partial x_i}\frac{\partial C(\mathbf{x})}{\partial x_j} \right\vert_{\mathbf{x}=0},
\]
for any real coordinate system $\mathbf{x}\in\mathbb{R}^N$. A detailed derivation is given in Appendix~\ref{appendix:Proof_of_Effective_Second_Order_Spectral_Moment}.

\subsection{\label{sec:Local_Maxima_Statistics}Local Maxima Statistics}
Despite the fact that intensity has a $\chi_2^2$ distribution, the distribution of speckles are not $\chi_2^2$ due to a bias in the sampling process. Counting speckles restricts our sampling process to the local maxima of intensity, and therefore Kac-Rice/local maxima theory must be used to determine their distribution \cite{adler2010geometry,adler2007random}. While the asymptotic local maxima theory for $\chi_2^2$ random fields is well established in mathematical literature, its application to laser speckle requires it to be written in a representation where the field gradient cross-correlations vanish. This is achieved by the phase shift in Eq.~\eqref{phase_shift}. We therefore specialize Worsley's \cite{worsley1994local} general result to the laser-speckle problem. It follows from \cite[Theorem 3.3]{worsley1994local} that, for a $\chi_2^2$ distributed random variable $I$, the expected number of local maxima above a normalized (with respect to the average) level set intensity $u:=I/I_0$ on a compact domain $\Omega\subset\mathbb{R}^N$ of Lebesgue measure $\lambda(\Omega)$ can be expressed asymptotically as
\begin{equation}\label{maxima_count_chi_2}
\langle M_\Omega^+(u) \rangle = \frac{\lambda(\Omega)\det(\mathring{\Lambda}^{(2)})^{\frac{1}{2}}}{\pi^{N/2}} u^{\frac{N}{2}}e^{-u}\left[1+\mathcal{O}(u^{-\frac{1}{2}})\right]
\end{equation}
where $\mathring{\Lambda}^{(2)}$ denotes the unit-normalized phase corrected second-order spectral moment (rank-2 tensor) of the component fields of $I$. The factor $\det(\mathring{\Lambda}^{(2)})^{\frac{1}{2}}$ gives the inverse local correlation volume associated with the field. In $N$ dimensions, let $\mathring{\lambda}_1,\mathring{\lambda}_2,\cdots,\mathring{\lambda}_N$ denote the eigenvalues of $\mathring{\Lambda}^{(2)}$. The corresponding principal correlation lengths are $\ell_j=\mathring{\lambda}_j^{-1/2}$, so that $\det(\mathring{\Lambda}^{(2)})^{\frac{1}{2}} = (\ell_1\cdots\ell_N)^{-1}$. These correlation lengths may be interpreted as the characteristic, or typical, speckle scales along the principal directions. In our calculations,
\begin{equation}\label{eq:correlation_length}
\mathring{\Lambda}^{(2)} = \mathrm{diag}(\ell_\perp^{-2},\ell_\perp^{-2},\ell_z^{-2},\ell_t^{-2}),
\end{equation}
where $\ell_\perp$ denotes the typical transverse speckle width, $\ell_z$ the typical longitudinal speckle length, and $\ell_t$ the typical speckle duration~\footnote{If the paraxial Eq.~\eqref{eq:paraxial_pde} were not transformed to retarded time, the matrix $\mathring{\Lambda}^{(2)}$ would contain off-diagonal entries. However, its determinant and eigenvalues would remain unchanged, thus the statistics would remain unchanged.}.

\begin{figure}
\includegraphics[width=0.49\linewidth]{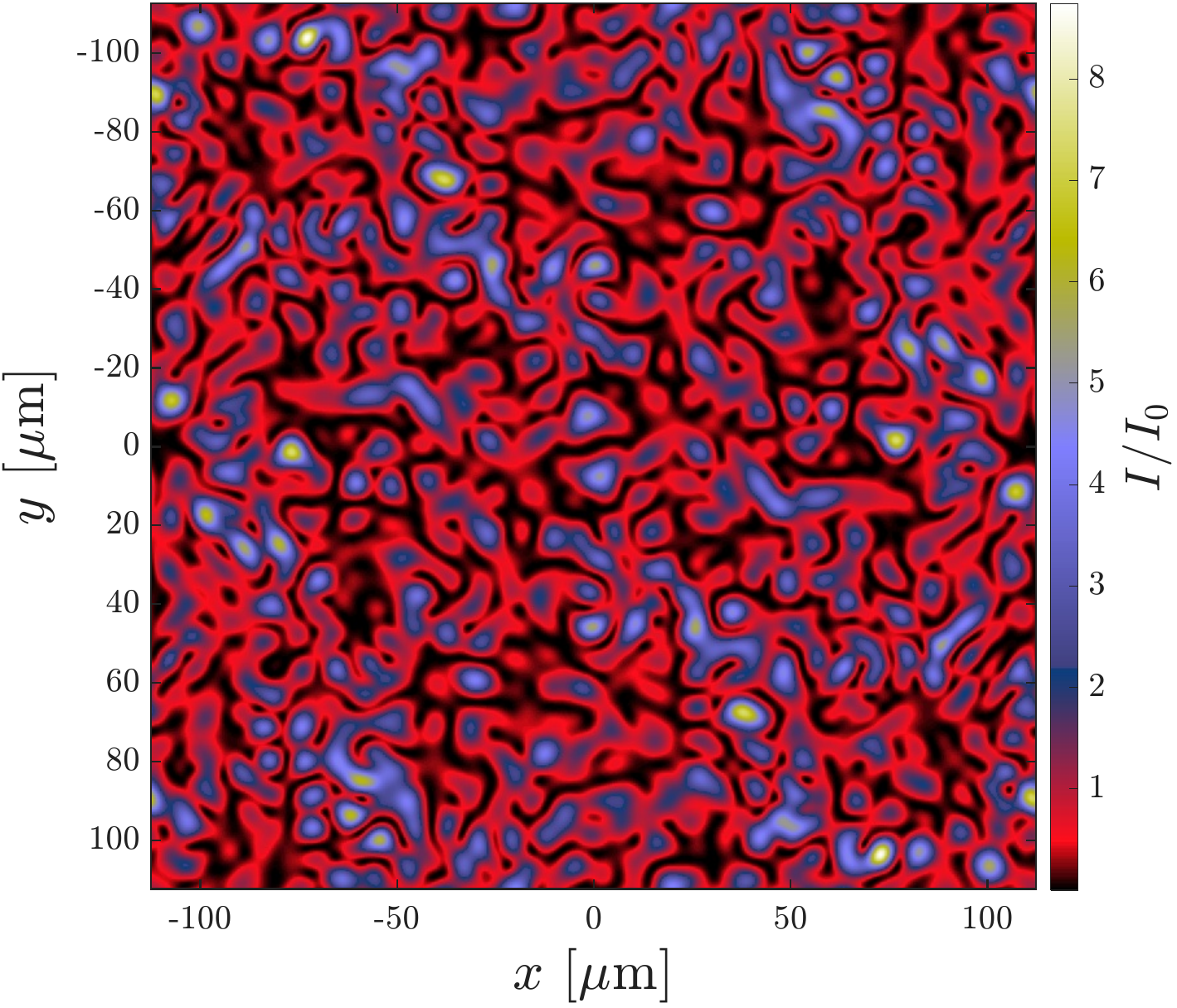}
\includegraphics[width=0.49\linewidth]{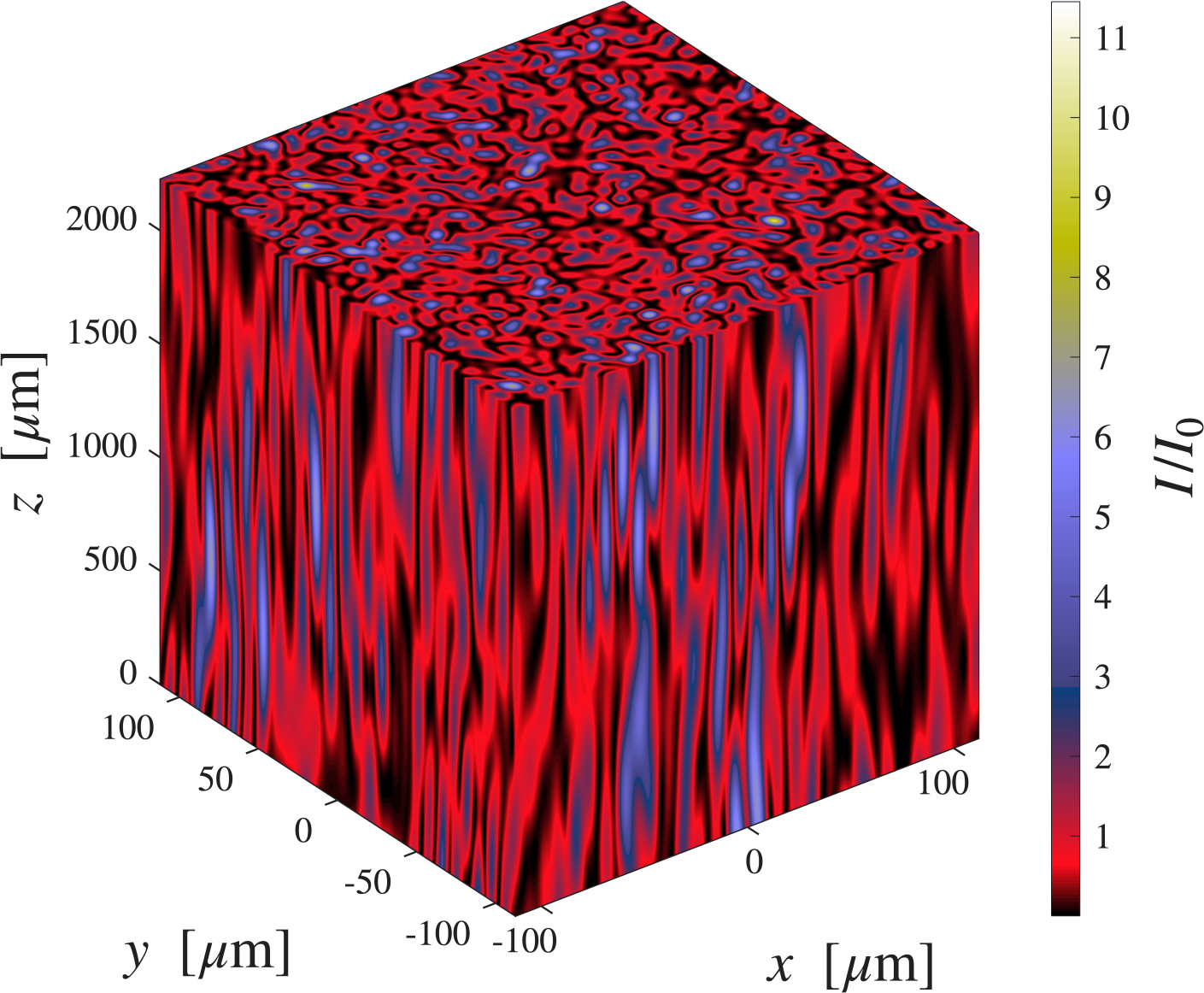}
\caption{Example realization of the normalized speckle intensity field due to RPP beams in real space in the $\mathbb{R}^2$ plane (left) and $\mathbb{R}^3$ volume (right).}
\label{fig:rpp_speckles}
\end{figure}

\subsection{\label{sec:Computation_of_Statistical_Quantities}Computation of Statistical Quantities}
Eq.~\eqref{maxima_count_chi_2} shows that the local maxima count is completely determined by the phase-corrected second-order spectral moments. We therefore compute these moments for several aperture geometries. Some of these geometries have been considered previously in the literature \cite{rose1993statistical,garnier1999statistics}. For consistency with the present formulation, we recompute the corresponding phase-corrected moments.

\subsubsection{Square Geometry}
The unit-normalized spectral density for a laser beam with a "top-hat" square envelope and Gaussian shaped pulse can be computed using equations~\eqref{boundary_square},~\eqref{boundary_temporal}, and~\eqref{eq:spectral_density_A}
\begin{equation}
S_{\square}=\frac{(2\pi)^3 \ell_t}{\sqrt{2\pi}\,(2k_c)^2}\,\mathbf{1}_{[-k_c,k_c]^2}\exp\left(-\frac{\ell_t^2\omega^2}{2}\right).
\end{equation}
Using Eq.~\eqref{eq:covariance_spectral}, we obtain
\begin{equation}\label{cov_square_rpp}
C_{\square} = \frac{1}{4k_{c}^2}\exp\left(-\frac{t^2}{2\ell_t^2}\right) \iint_{-k_{c}}^{k_{c}} e^{i\mathbf{k}_\perp \cdot \mathbf{x}_\perp-\frac{ik_\perp^2 z}{2k_0}}\, d^2\mathbf{k}_\perp.
\end{equation}
This implies that the phase corrected second-order spectral moment (for $\mathbb{R}^4$) can be calculated as
\begin{equation}\label{second_order_spectral_moment_square_matrix_form}
\mathring{\Lambda}_{\square}^{(2)} = \mathrm{diag}\left(\frac{k_{c}^2}{3},\ \frac{k_{c}^2}{3},\ \frac{2k_{c}^4}{45k_0^2},\ \frac{1}{\ell_t^2}\right).
\end{equation}

\subsubsection{Circular Geometry}
The unit-normalized spectral density for a laser beam with a "top-hat" circular envelope and Gaussian shaped pulse can be computed using equations~\eqref{boundary_circular},~\eqref{boundary_temporal}, and~\eqref{eq:spectral_density_A}
\begin{equation}
S_{\circ} = \frac{(2\pi)^3 \ell_t}{\sqrt{2\pi}\,\pi k_c^2}\,\mathbf{1}_{\{|\mathbf{k}_{\perp}|\leq k_c\}}\exp\left(-\frac{\ell_t^2\omega^2}{2}\right).
\end{equation}
Using Eq.~\eqref{eq:covariance_spectral}, we obtain
\begin{equation}\label{cov_circular_rpp}
C_{\circ} = \frac{2}{k_{c}^2}\exp\left(-\frac{t^2}{2\ell_t^2}\right)\int_{0}^{k_{c}} k_\perp J_0(k_\perp x_\perp)\,e^{\frac{k_\perp^2 z}{2ik_0}}  \,dk_\perp,
\end{equation}
where $x_\perp = |\mathbf{x}_\perp|$, $k_\perp = |\mathbf{k}_\perp|$, and $J_0$ is the Bessel function of the first kind of zeroth order. This implies the phase corrected second-order spectral moment can be calculated as
\begin{equation}\label{second_order_spectral_moment_circular_matrix_form}
\mathring{\Lambda}_{\circ}^{(2)} = \mathrm{diag}\left(\frac{k_{c}^2}{4},\ \frac{k_{c}^2}{4},\ \frac{k_{c}^4}{48k_0^2},\ \frac{1}{\ell_t^2}\right).
\end{equation}

\subsubsection{Annular Geometry}
The unit-normalized spectral density for a laser beam with a "top-hat" annular envelope and Gaussian shaped pulse can be computed using equations~\eqref{boundary_annular},~\eqref{boundary_temporal}, and~\eqref{eq:spectral_density_A}
\begin{equation}
S_{\circledcirc}=\frac{(2\pi)^3 \ell_t}{\sqrt{2\pi}}\,\frac{\mathbf{1}_{\{k_I\leq|\mathbf{k}_{\perp}|\leq k_O\}}}{\pi(k_O^2-k_I^2)}\exp\left(-\frac{\ell_t^2\omega^2}{2}\right).
\end{equation}
Using Eq.~\eqref{eq:covariance_spectral}, we obtain
\begin{equation}\label{cov_annular_rpp}
C_{\circledcirc} = \frac{2\exp\left(-\frac{t^2}{2\ell_t^2}\right)}{k_O^2-k_I^2}\int_{k_I}^{k_O} k_\perp J_0(k_\perp x_\perp)\,e^{\frac{k_\perp^2 z}{2ik_0}}  \,dk_\perp.
\end{equation}
This implies the phase corrected second-order spectral moment can be calculated as
\begin{equation}\label{second_order_spectral_moment_annular_matrix_form}
\mathring{\Lambda}_{\circledcirc}^{(2)} = \mathrm{diag}\left(\frac{k_O^2+k_I^2}{4}, \frac{k_O^2+k_I^2}{4}, \frac{(k_O^2-k_I^2)^2}{48k_0^2}, \frac{1}{\ell_t^2}\right).
\end{equation}

\subsubsection{Gaussian Geometry}
The unit-normalized spectral density for a laser beam with a Gaussian envelope and Gaussian shaped pulse can be computed using equations~\eqref{boundary_gaussian},~\eqref{boundary_temporal}, and~\eqref{eq:spectral_density_A}
\begin{equation}
S_g=(2\pi)^{3/2} L_{\perp}^2 \ell_t\,\exp\left(-\frac{L_{\perp}^2|\mathbf{k}_{\perp}|^2}{2}-\frac{\ell_t^2\omega^2}{2}\right).
\end{equation}
Using Eq.~\eqref{eq:covariance_spectral}, we obtain
\begin{equation}\label{cov_Gaussian_rpp}
C_{g} = \frac{L_{\perp}^2}{L_{\perp}^2+i\frac{z}{k_0}}  \exp\left(-\frac{x_\perp^2}{2\left(L_{\perp}^2+i\frac{z}{k_0}\right)}-\frac{t^2}{2\ell_t^2}\right).
\end{equation}
This implies the phase corrected second-order spectral moment can be calculated as
\begin{equation}\label{second_order_spectral_moment_gaussian_matrix_form}
\mathring{\Lambda}_{g}^{(2)} = \mathrm{diag}\left(\frac{1}{L_{\perp}^2},\ \frac{1}{L_{\perp}^2},\ \frac{1}{L_{\perp}^4k_0^2},\ \frac{1}{\ell_t^2}\right).
\end{equation}

\begin{figure}
\includegraphics[width=\linewidth]{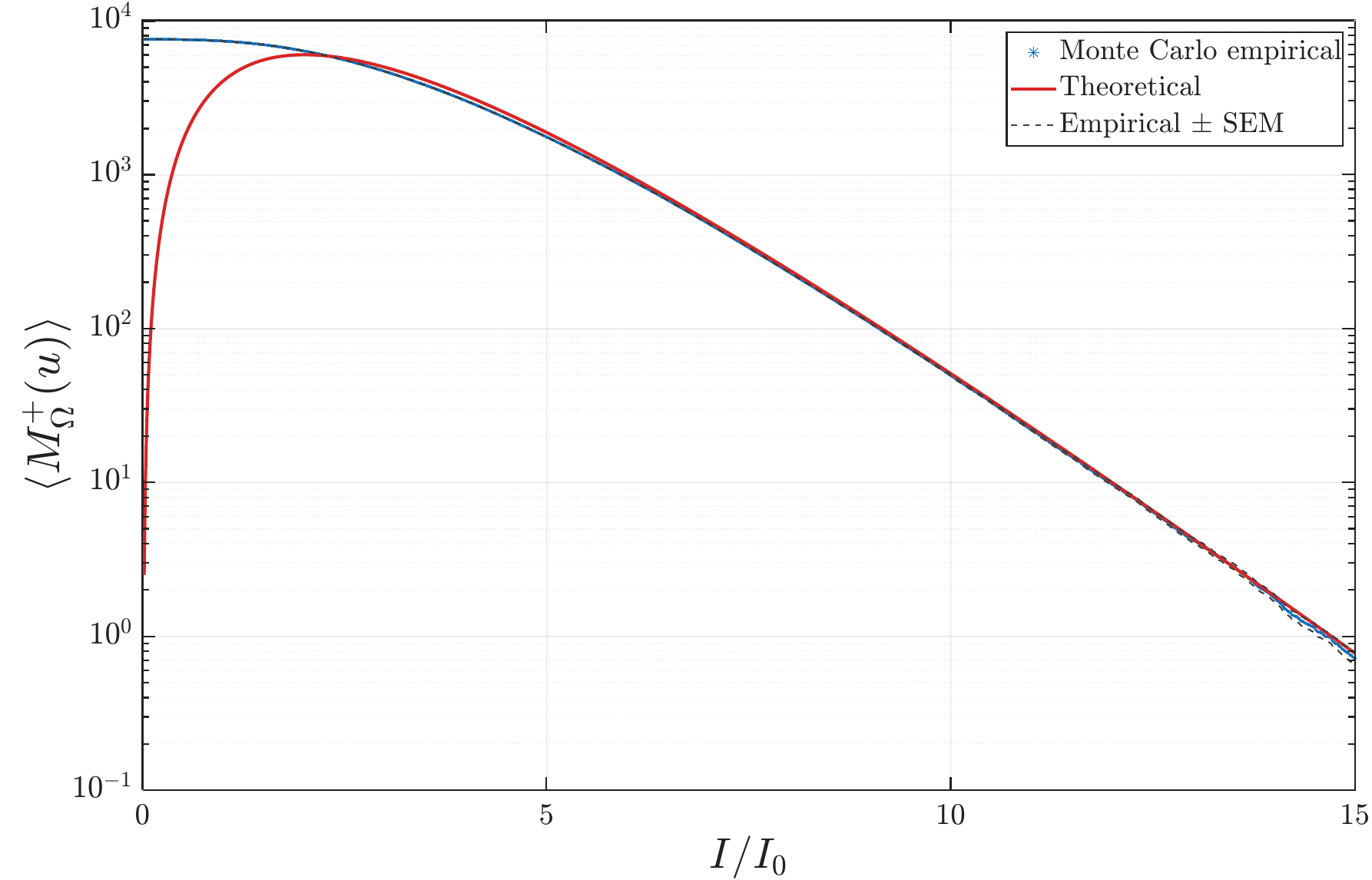}
\caption{Monte Carlo comparison between empirical and theoretical distributions of local maxima intensities for a square aperture geometry with domain $(\mathbf{x}_{\perp},z,t)\in\mathbb{R}^4$. Simulations use 200 realizations of a square RPP with $\tilde{f}/\tilde{D}=20$, a $3\omega$ beam, $N_{\mathrm{RPP}}=16$ and a $\ell_t$ which is 25 times shorter than the pulse duration. Distributions are shown on a logarithmic scale.}
\label{fig:speckle_distribution_geometries_xyzt}
\end{figure}

\begin{table}
\renewcommand{\arraystretch}{2}
\begin{ruledtabular}
\begin{tabular}{ccccc}
 & $\mathbf{x}_{\perp}$ & $(x_{j},z)$ & $(\mathbf{x}_{\perp},z)$ & $(\mathbf{x}_{\perp},z,t)$ \\
\hline
Dimensions 
& $\mathbb{R}^{2}$ 
& $\mathbb{R}^{2}$ 
& $\mathbb{R}^{3}$ 
& $\mathbb{R}^{4}$ \\
\hline
Square 
& $\frac{k_{c}^2}{3}$ 
& $\frac{2k_{c}^3}{3\sqrt{30}\,k_0}$ 
& $\frac{2k_{c}^4}{9\sqrt{10}\,k_0}$ 
& $\frac{2k_{c}^4}{9\sqrt{10}\,\ell_tk_0}$ \\
Circular 
& $\frac{k_{c}^2}{4}$ 
& $\frac{k_{c}^3}{8\sqrt{3}k_0}$ 
& $\frac{k_{c}^4}{16\sqrt{3}\,k_0}$ 
& $\frac{k_{c}^4}{16\sqrt{3}\,\ell_tk_0}$ \\
Annular 
& $\frac{k_O^2+k_I^2}{4}$ 
& $\frac{\sqrt{k_O^2+k_I^2}(k_O^2-k_I^2)}{8\sqrt{3}k_0}$ 
& $\frac{k_O^4-k_I^4}{16\sqrt{3}\,k_0}$ 
& $\frac{k_O^4-k_I^4}{16\sqrt{3}\,\ell_tk_0}$ \\
Gaussian 
& $\frac{1}{L_{\perp}^2}$ 
& $\frac{1}{L_{\perp}^3k_0}$ 
& $\frac{1}{L_{\perp}^4k_0}$ 
& $\frac{1}{L_{\perp}^4\ell_tk_0}$ \\
\end{tabular}
\end{ruledtabular}
\caption{Determinants of the unit-normalized second-order spectral moment matrices, $\det(\mathring{\Lambda}^{(2)})^{1/2}$, for square, circular, annular, and Gaussian apertures, over different domains. Note $x_j$ here is a single component of $\mathbf{x}_\perp$ (either component is equivalent by symmetry).}
\label{table_aperture_local_maxima_determinants}
\end{table}
For each entry in Table~\ref{table_aperture_local_maxima_determinants}, the corresponding expected number intensity local maxima above level set $u$, up to leading order, is
\begin{align}
\mathbb{R}^2: & \qquad \langle M_\Omega^+(u) \rangle = \lambda(\Omega)\det(\mathring{\Lambda}^{(2)})^{\frac{1}{2}} \frac{u}{\pi} e^{-u},\label{my_Counting_Formula_2}\\
\mathbb{R}^3: & \qquad \langle M_\Omega^+(u) \rangle = \lambda(\Omega)\det(\mathring{\Lambda}^{(2)})^{\frac{1}{2}} \left(\frac{u}{\pi}\right)^{\frac{3}{2}}e^{-u},\label{my_Counting_Formula_3}\\
\mathbb{R}^4: & \qquad \langle M_\Omega^+(u) \rangle = \lambda(\Omega)\det(\mathring{\Lambda}^{(2)})^{\frac{1}{2}} \left(\frac{u}{\pi}\right)^{2}e^{-u}.\label{my_Counting_Formula_4}
\end{align}
for each dimension of the domain under consideration. For example Fig.~\ref{fig:speckle_distribution_geometries_xyzt} uses Eq.~\eqref{my_Counting_Formula_4}.

\subsection{\label{sec:single_component_ansatz}Comparison with the Single Component Ansatz}
\begin{figure*}
\includegraphics[width=0.49\linewidth]{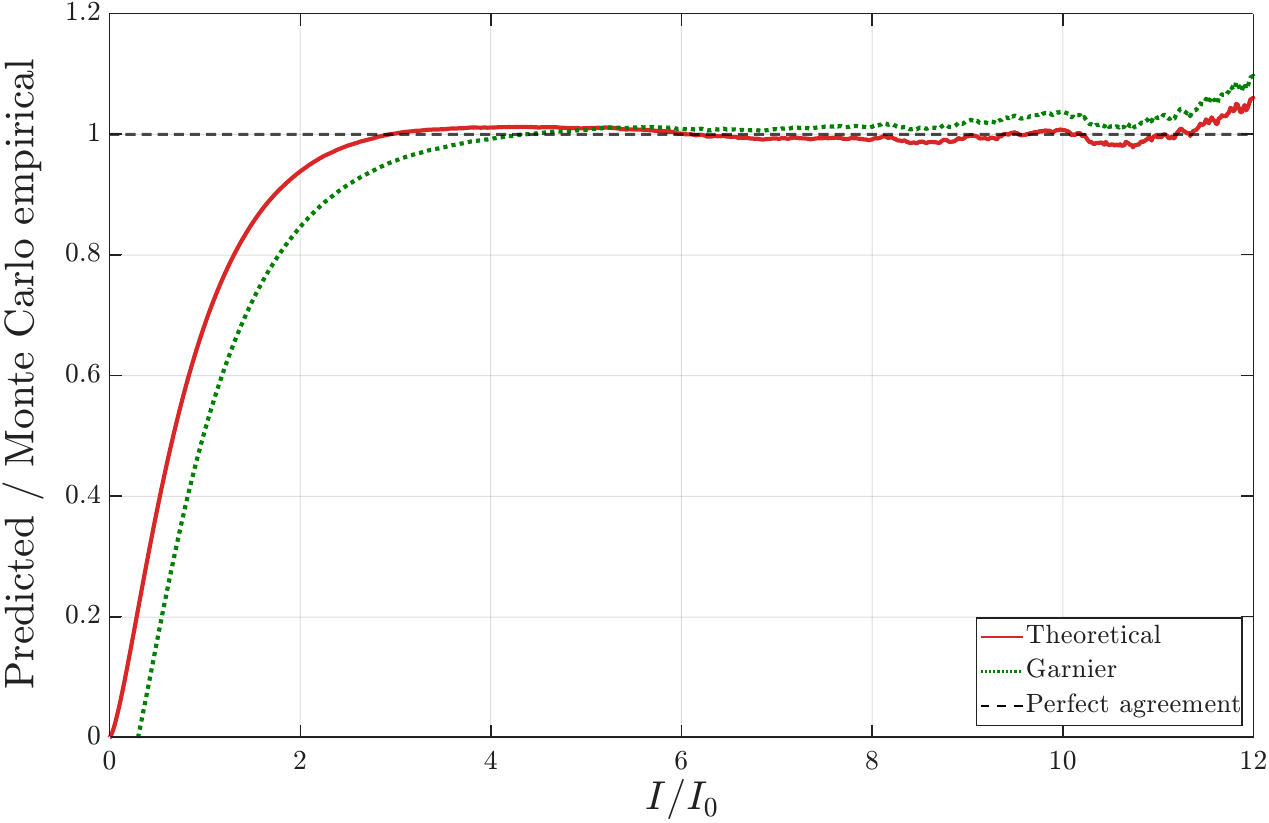}
\hfill
\includegraphics[width=0.49\linewidth]{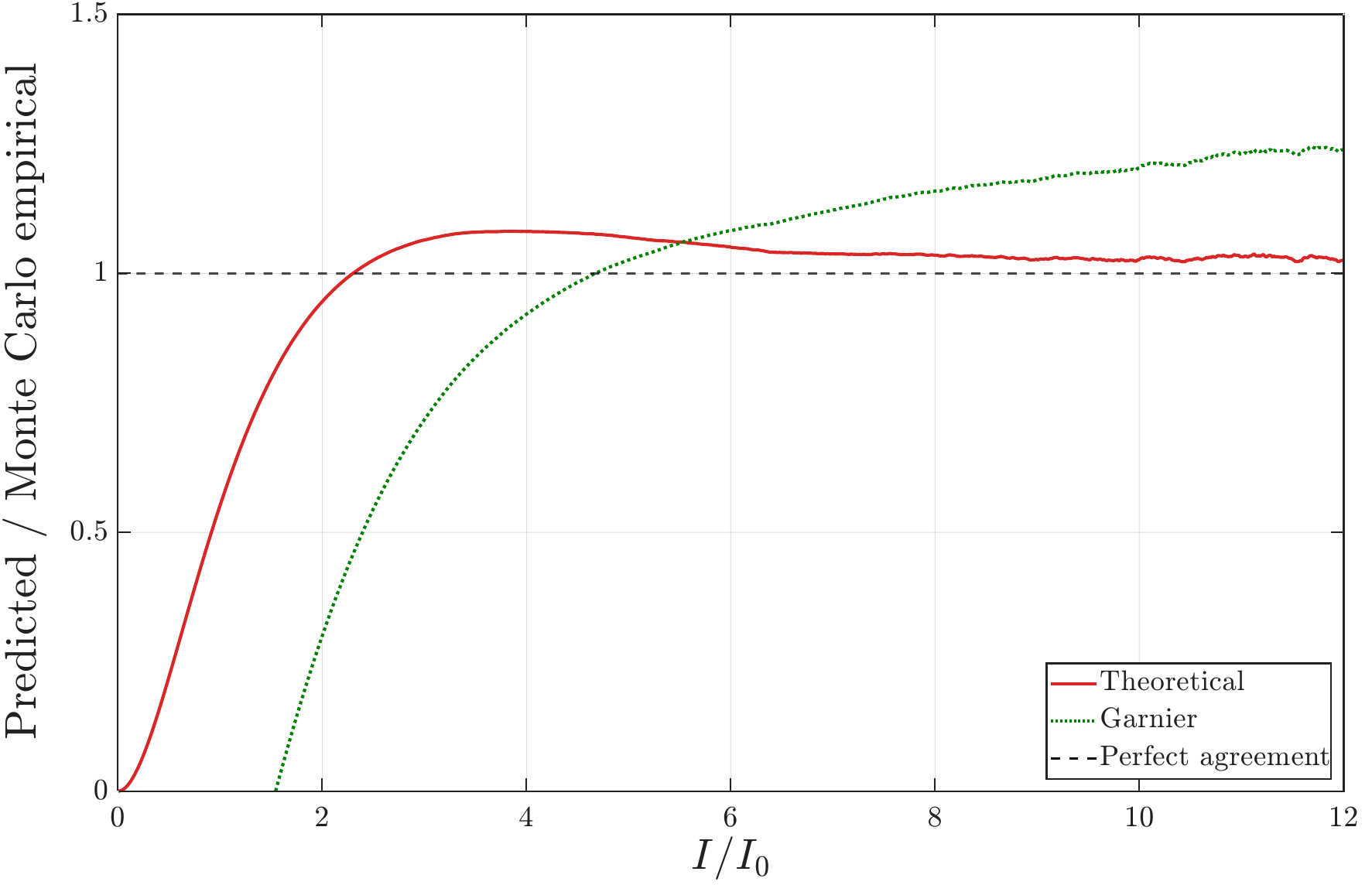}
\caption{Monte Carlo comparison of local maxima count predictions for square RPP speckle fields. Results are shown comparing our theory and that of Garnier \cite{garnier1999statistics} in both $(\mathbf{x}_{\perp},z)\in\mathbb{R}^3$ (left) and $(\mathbf{x}_{\perp},z,t)\in\mathbb{R}^4$ with ISI (right). The plotted quantity is the ratio of theoretical to numerical local maxima counts. Simulations use 200 realizations of a square RPP with $\tilde{f}/\tilde{D}=20$ and a $3\omega$ beam, with $N_{\mathrm{RPP}}=64$ and $N_{\mathrm{RPP}}=16$ for the $\mathbb{R}^3$ and $\mathbb{R}^4$ cases, respectively.}
\label{fig:ratio_garnier}
\end{figure*}
Previous methods for calculating the expected number of speckles above an intensity level set were developed by Rose \& DuBois \cite{rose1993statistical} and Garnier \cite{garnier1999statistics}. Both approaches rely on a single component ansatz, by which the local maxima of the intensity field, $I \propto \mathcal{A}_R^2+\mathcal{A}_I^2$, are identified with the local maxima of either the real $\mathcal{A}_R^2$ or imaginary $\mathcal{A}_I^2$ component of the underlying complex field. Specifically, Rose \& DuBois's ansatz had relied on the fact that for a speckle associated with the local maxima of $\mathcal{A}_R^2$ (or $\mathcal{A}_I^2$), i.e. when $\nabla \mathcal{A}_R^2 = 0$ (or $\nabla \mathcal{A}_I^2 = 0$), the contributions of $\mathcal{A}_I^2$ (or $\mathcal{A}_R^2$) were negligible. Garnier's correction states that, for speckles associated with local maxima of $\mathcal{A}_R^2$ (or $\mathcal{A}_I^2$), the contribution of $\mathcal{A}_I^2$ (or $\mathcal{A}_R^2$) is modelled as an independent scaled $\chi^2_N$ contribution to the intensity in $N$ field dimensions. However, a true local maximum of the intensity satisfies
\begin{equation}
\nabla I=0 \quad \implies \quad \nabla \mathcal{A}_R^2+\nabla \mathcal{A}_I^2=0,
\label{eq:intensity_gradient_condition}
\end{equation}
which permits cancellation between the component gradients and therefore does not require either gradient to vanish individually. While the single component ansatz has served as the foundation of existing speckle counting theories, the intensity gradient conditix~\eqref{eq:intensity_gradient_condition}, motivates comparison with a theory based directly on the maxima of the full intensity field. In Fig.~\ref{fig:ratio_garnier}, we show the comparison of Garnier's single component ansatz (corrected from Rose \& DuBois's original formulation) and our presented theory against numerical Monte Carlo simulations of speckles. The discrepancy introduced by the single component ansatz becomes apparent, particularly in $\mathbb{R}^4$. Garnier's formula (under our covariance conventions) is given in Appendix~\ref{appendix:Garniers_Counting_Formula}, while the formulas derived in this work are given in equations~\eqref{my_Counting_Formula_3} and~\eqref{my_Counting_Formula_4}.

\section{Applications}
We now apply the speckle statistics developed above in three settings. First, we use the expected number of speckle maxima to construct a simple model for reflectivity saturation. Second, we compare speckle-driven density fluctuations with thermal density fluctuations through their spectral densities. Lastly, we discuss how speckle sizes should be interpreted, and in particular how this interpretation depends on the statistical quantity used to define a characteristic length.

\subsection{\label{sec:Reflectivity_Saturation_of_an_RPP_Beam}Reflectivity Saturation of an RPP Beam}
First introduced in~\cite{rose1994laser}, and later derived in~\cite{michel2023introduction} using Garnier's~\cite{garnier1999statistics} speckle formula, a simple model for the average stimulated Brillouin scattering (SBS) reflectivity generated by intense speckles in a plasma of volume $\lambda(\Omega)$ can be formulated by assuming that only speckles whose normalized intensities exceed a threshold value $u_{\mathrm{th}}$ contribute to the reflectivity. This threshold-based model is motivated by the fact that linear SBS estimates based on the average laser intensity can miss the contribution of rare, intense speckles, whose local gain may be large even when the mean intensity remains below the nonlinear threshold. The corresponding average reflectivity is then
\begin{equation}\label{eq:reflectivity}
\langle R \rangle = - \int_{u_{\mathrm{th}}}^{\infty} P(u)\, dM_\Omega^+(u),
\end{equation}
where $-dM_\Omega^+(u)$ is the expected number of local maxima whose intensities lie in the infinitesimal interval $[u,u+du]$, and $P(u) = \frac{\pi\ell_{\perp}^2u}{\lambda(\Omega_{\perp})}$ is the fraction of the total beam power contained in a speckle where $\Omega = \Omega_{\perp}\times\Omega_{z}$. Thus using the 3 dimensional speckle count, Eq.~\eqref{my_Counting_Formula_3}, and using Eq.~\eqref{eq:correlation_length} with $\lambda(\Omega) = \lambda(\Omega_{\perp})\lambda(\Omega_{z})$, we get that 
\begin{equation}
\langle R \rangle = \frac{\lambda(\Omega_z)}{\sqrt{\pi}\ell_z} \int_{u_{\mathrm{th}}}^{\infty} \left(u^{5/2}-\frac{3}{2}u^{3/2} \right)e^{-u}\, du.
\end{equation}
To evaluate the integral, we apply Laplace's method for endpoint asymptotics for $u_{\mathrm{th}}\gg 1$, so that up to leading order, we arrive at
\begin{equation}
\langle R \rangle = \frac{1}{\sqrt{\pi}}\frac{\lambda(\Omega_z)}{\ell_z} u_{\mathrm{th}}^{5/2}e^{-u_{\mathrm{th}}}\left[1+\mathcal{O}(u_{\mathrm{th}}^{-1})\right].
\label{eq:reflectivity_formulation}
\end{equation}
While Eq.~\eqref{eq:reflectivity_formulation} exhibits the same leading-order asymptotic dependence as Eq.~(9.121) of Michel~\cite{michel2023introduction}, the prefactor differs because Ref.~\cite{michel2023introduction} employs Garnier's single-component counting theory and the conventional longitudinal speckle length, whereas the present work uses a three-dimensional $\chi_2^2$ local maxima count together with speckle dimensions defined by the second-order spectral moments. The difference in prefactors therefore reflects both the underlying counting theory and the adopted definition of speckle size. We simulate the average reflectivity by generating Monte Carlo realizations of the speckle field and replacing the integral in Eq.~\eqref{eq:reflectivity} with the corresponding discrete sum over local maxima, namely for $N_{\mathrm{run}}$ realizations
\begin{equation}
\langle R_{\mathrm{sim}} \rangle = \frac{1}{N_{\mathrm{run}}}\sum_{n=1}^{N_{\mathrm{run}}} \sum_{j:u_j^{(n)}>u_{\mathrm{th}}} P(u_j^{(n)}).
\label{eq:numerical_reflectivity_formulation}
\end{equation}
We show that the present theory reproduces both the asymptotic shape and the numerical pre-factor of the simulated reflectivity.
\begin{figure}
\includegraphics[width=\linewidth]{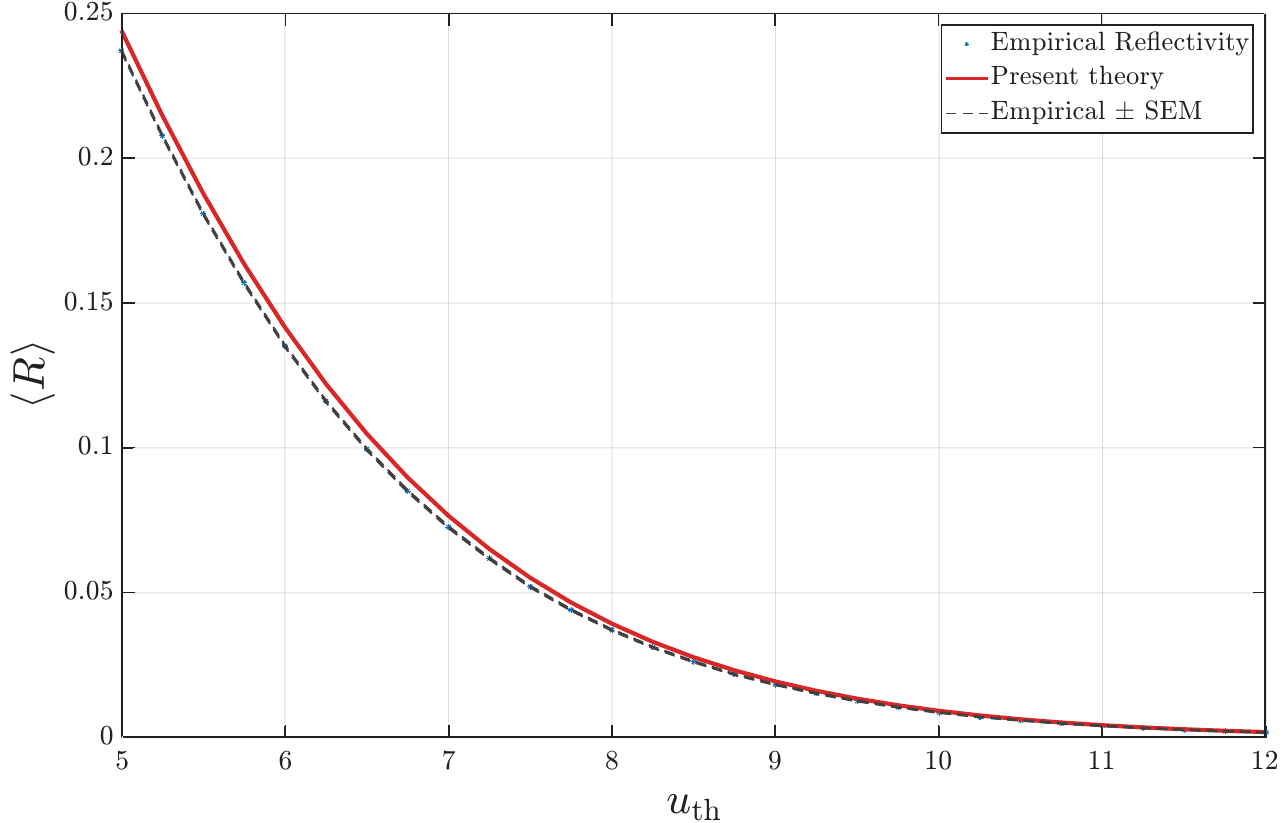}
\caption{Comparison of theoretical (Eq.~\eqref{eq:reflectivity_formulation}) and numerical (Eq.~\eqref{eq:numerical_reflectivity_formulation}) average reflectivity for a circular aperture as a function of the normalized intensity threshold $u_{\mathrm{th}}\in[5,12]$, using 500 speckle field realizations. The threshold is sampled in increments of $0.25$. The asymptotic theory captures the correct overall scale of the numerical average reflectivity, including the prefactor.}
\label{fig:reflectivity_xyz}
\end{figure}

\subsection{\label{sec:Density_Fluctuations}Density Fluctuations}
\begin{table*}
\begin{ruledtabular}
\begin{tabular}{cccc}
 & $\mathbf{x}_{\perp}\in\mathbb{R}^2$ & $(x_{j},z)\in\mathbb{R}^2$ & $(\mathbf{x}_{\perp},z)\in\mathbb{R}^3$ \\
\hline
Square 
& $L_{\perp}^2=\frac{3}{k_c^2}$ 
& $L_{\perp}^2=\left(\frac{135}{2}\right)^{\frac{1}{3}}\frac{1}{k_c^2}$  
& $L_{\perp}^2=\left(\frac{405}{2}\right)^{\frac{1}{4}}\frac{1}{k_c^2}$  \\
Circular 
& $L_{\perp}^2=\frac{4}{k_c^2}$ 
& $L_{\perp}^2=(192)^{\frac{1}{3}}\frac{1}{k_c^2}$ 
& $L_{\perp}^2=(768)^{\frac{1}{4}}\frac{1}{k_c^2}$ \\
Annular
& $L_{\perp}^2=\frac{4}{k_O^2+k_I^2}$ 
& $L_{\perp}^2=\left(\frac{192}{(k_O^2+k_I^2)(k_O^2-k_I^2)^2}\right)^{\frac{1}{3}}$
& $L_{\perp}^2=\left(768\right)^{\frac{1}{4}}\frac{1}{\sqrt{k_O^4-k_I^4}}$ \\
\end{tabular}
\end{ruledtabular}
\caption{Gaussian transverse width $L_{\perp}$ chosen to match the speckle statistics for square, circular, and annular top-hat spectra in different spatial sections. For four-dimensional $(\mathbf{x}_{\perp},z,t)\in\mathbb{R}^4$ statistics, the three-dimensional value associated with $(\mathbf{x}_{\perp},z)\in\mathbb{R}^3$ is used.}
\label{table_gaussian_beam_width}
\end{table*}

Stable laser-produced plasmas can sustain enhanced levels of electron density fluctuations. Such non-equilibrium states may arise in the sub-threshold regime of parametric instabilities \cite{oberman1974general,berger1989effect,carleton2026geometric}, or be generated by the stochastic ponderomotive force \cite{brantov1999plasma, grech2006plasma} and subsequently amplified through the thermal plasma response \cite{brantov1999plasma} to a randomized laser pulse.

In this subsection, we apply our theoretical model to calculate the spectra of electron density fluctuations and compare them with the equilibrium fluctuations arising from particle discreteness under the same plasma conditions. The latter constitute the standard fluctuation background commonly used in the interpretation of Thomson scattering experiments \cite{froula2010plasma}.\\

We define the spectral density due to density fluctuations as $S_{\delta n_e\delta n_e}\colon\mathbb{R}^4\to\mathbb{C}$ though Eq.~\eqref{spectral_density_definition} in Appendix~\ref{sec:Some_Math_Definitions}, namely for $(\mathbf{k},\omega)\in\mathbb{R}^4$
\begin{eqnarray}
&& \left\langle \delta \hat{n}_e(\mathbf{k},\omega)\overline{\delta \hat{n}_e(\mathbf{k}',\omega')}\right\rangle \nonumber\\
&& = (2\pi)^N S_{\delta n_e\delta n_e}(\mathbf{k}_\perp,\omega)\delta(\mathbf{k}_\perp-\mathbf{k}_\perp')\delta(\omega-\omega').
\end{eqnarray}
Let $s\in\{e,i\}$ define a species index, then let $n_{s0}$ denote the equilibrium density, $T_s$ denote the temperature in energy units, and $n_c=\varepsilon_0m_e\omega_0^2/q_e^2$ denote the critical electron density, then the spectral density of $\delta n_e$ due to speckles can be written as
\begin{equation}\label{eq:density_fluctuations}
S_{\delta n_e\delta n_e}\left(\mathbf{k},\omega\right) = \frac{n_{e0}^2}{4c^2T_e^2n_c^2} \left\lvert \hat{\mathcal{H}}(\mathbf{k},\omega) \right\rvert^2 S_{II}\left(\mathbf{k},\omega\right),
\end{equation}
where $S_{II}$ is the spectral density defined though~\eqref{spectral_density_definition} in Appendix~\ref{sec:Some_Math_Definitions}, namely
\begin{eqnarray}
&& \left\langle \hat{I}(\mathbf{k},\omega)\overline{\hat{I}(\mathbf{k}',\omega')}\right\rangle \nonumber\\
&& = (2\pi)^N S_{II}(\mathbf{k}_\perp,\omega)\delta(\mathbf{k}_\perp-\mathbf{k}_\perp')\delta(\omega-\omega').
\end{eqnarray}
and where $\hat{\mathcal{H}}$ denotes the response function in Fourier space. $\hat{\mathcal{H}}$ can be written for either kinetic theory derived in Appendix~\ref{Derivation_of_Kinetic_Density_Response_Due_to_Speckles}, or fluid theory shown in \cite{brantov1999plasma}, which are respectively expressed as
\begin{align}
\hat{\mathcal{H}}(\mathbf{k},\omega) &= -k^2\lambda_{De}^2\frac{1+\chi_i(\mathbf{k},\omega)}{\varepsilon(\mathbf{k},\omega)}\chi_e(\mathbf{k},\omega),\label{eq:kinetic_response_function_kw} \\
\hat{\mathcal{H}}(\mathbf{k},\omega) &= \left[\left(\frac{\omega}{kc_s}\right)^2+2i\frac{\gamma_a\omega}{kc_s^2}-1\right]^{-1},
\end{align}
where $\lambda_{Ds}^2=\varepsilon_0T_s/(n_{s0}q_s^2)$ is the Debye length squared for a species $s$, $c_s$ is the ion acoustic speed, $\gamma_a$ is the acoustic damping coefficient, $\chi_s(\mathbf{k},\omega)$ is the susceptibility for a species $s$, and $\varepsilon(\mathbf{k},\omega)$ is the permittivity. We can then find the spectral density of intensity through the covariance of $\mathcal{A}$ as follows. 

The intensity has a non-zero mean, i.e. $\langle I\rangle = I_0$, so we define its covariance as
\begin{equation}
C_{II}(\mathbf{x},\mathbf{x}^{\prime}) = \left\langle I(\mathbf{x})I(\mathbf{x}^\prime)\right\rangle - \langle I\rangle^2.
\end{equation}
Substituting Eq.~\eqref{intensity_definition}, we get
\begin{equation}\label{cov_intensity}
C_{II}(\mathbf{x},\mathbf{x}^{\prime}) = \frac{\varepsilon_0^2 c^2\, \omega_0^4}{4} \left[ \left\langle |\mathcal{A}(\mathbf{x})|^2 |\mathcal{A}(\mathbf{x}^{\prime})|^2 \right\rangle - \left\langle |\mathcal{A}|^2\right\rangle^2 \right].
\end{equation}
Since $\mathcal{A}$ is a proper complex Gaussian field, Wick's probability theorem gives
\begin{equation}\label{cov_A_squared}
\left\langle |\mathcal{A}(\mathbf{x})|^2 |\mathcal{A}(\mathbf{x}^\prime)|^2 \right\rangle = \left\langle |\mathcal{A}|^2\right\rangle^2 + |C_{\mathcal{A}\mathcal{A}}(\mathbf{x},\mathbf{x}^{\prime})|^2.
\end{equation}
Therefore, substituting Eq.~\eqref{cov_A_squared} into Eq.~\eqref{cov_intensity} and recalling that $\mathcal{A}$ is homogeneous ($C_{\mathcal{A}\mathcal{A}}$ only depends on one variable), we get that
\begin{equation}
C_{II}(\mathbf{x})=\frac{\varepsilon_0^2 c^2\, \omega_0^4}{4} |C_{\mathcal{A}\mathcal{A}}(\mathbf{x})|^2.
\end{equation}
Taking the Fourier transform $\mathcal{F}:L^2(\mathbb{R}^4)\to L^2(\mathbb{R}^4)$ defined in Appendix~\ref{sec:Some_Math_Definitions} for 4 dimensions, with $(\mathbf{x},t)\mapsto (\mathbf{k},\omega)$ gets us
\begin{equation}
S_{II}(\mathbf{k},\omega) = \frac{\varepsilon_0^2 c^2\, \omega_0^4}{4} \mathcal{F}\left[|C_{\mathcal{A}\mathcal{A}}|^2\right].
\end{equation}
We will use the Gaussian aperture covariance to find a closed form expression for intensity spectral density, then to mimic the speckle statistics of physical configurations, we relate the beam width $L_{\perp}$ to physical parameters such as $\tilde{f}/\tilde{D}$ and $k_0$ through $k_c$ by matching the corresponding second-order spectral moments. Table~\ref{table_gaussian_beam_width} shows values of $L_{\perp}$ that reproduce the speckle statistics of our other configurations considered.

So, using Eq.~\eqref{cov_Gaussian_rpp} normalized to $\sigma_{\mathcal{A}\mathcal{A}}^2$, and shifting back to lab time from retarded time, which in Fourier space results to $k_z\mapsto k_z-\omega/v_g$, we get that the spectral density of intensity is
\begin{eqnarray}
S_{II}(\mathbf{k}_\perp,k_z,\omega) &=& I_0^2\frac{2\pi^2k_0 L_{\perp}^3 \ell_t}{|\mathbf{k}_\perp|} \exp\left( -\frac{L_{\perp}^2|\mathbf{k}_\perp|^2}{4}\right.\nonumber \\
 & &\left. -\frac{L_{\perp}^2k_0^2(k_z-\omega/v_g)^2}{|\mathbf{k}_\perp|^2} -\frac{\ell_t^2\omega^2}{4} \right),
\end{eqnarray}
where $(\mathbf{k}_\perp,k_z)=\mathbf{k}$. For an RPP beam without ISI, the temporal coherence time is effectively infinite on the timescales of interest. Equivalently, we assume $\ell_t |\omega| \gg 1$ except in an asymptotically narrow neighbourhood of $\omega=0$. The no-ISI RPP spectrum is therefore understood as the distributional limit $\ell_t\to\infty$, for which
\[
\frac{\ell_t}{2}\exp\left(-\frac{\ell_t^2\omega^2}{4}\right)
\to \sqrt{\pi}\delta(\omega).
\]
Upon integrating over frequency in the Fourier measure, 
\begin{equation}
\mathcal{S}_{\delta n_e\delta n_e}^{\mathrm{RPP}}(\mathbf{k}) := \frac{1}{2\pi}\int_{\mathbb{R}} S_{\delta n_e\delta n_e}(\mathbf{k},\omega)\,d\omega,
\end{equation}
the no-ISI limit yields the static density-fluctuation spectrum
\begin{align}
\mathcal{S}_{\delta n_e\delta n_e}^{\mathrm{RPP}}(\mathbf{k}) =& \frac{n_{e0}^2I_0^2}{2c^2T_e^2n_c^2}\frac{\pi^{3/2}k_0 L_{\perp}^3}{|\mathbf{k}_\perp|} \left\lvert \hat{\mathcal{H}}(\mathbf{k},0) \right\rvert^2\nonumber \\
& \times \exp\left( -\frac{L_{\perp}^2|\mathbf{k}_\perp|^2}{4} -\frac{L_{\perp}^2k_0^2k_z^2}{|\mathbf{k}_\perp|^2} \right),
\label{eq:density_fluctuations_frequency_integrated}
\end{align}
where
\begin{align}
\left\lvert \hat{\mathcal{H}}(\mathbf{k},0)\right\rvert^2 &= \left(\frac{k^2+\lambda_{Di}^{-2}}{k^2+\lambda_{De}^{-2}+\lambda_{Di}^{-2}}\right)^2,\label{eq:kinetic_response_function_k} \\
\left\lvert \hat{\mathcal{H}}(\mathbf{k},0)\right\rvert^2 &= 1,
\end{align}
for kinetic and fluid description respectively.
\begin{figure*}
\includegraphics[width=0.49\linewidth]{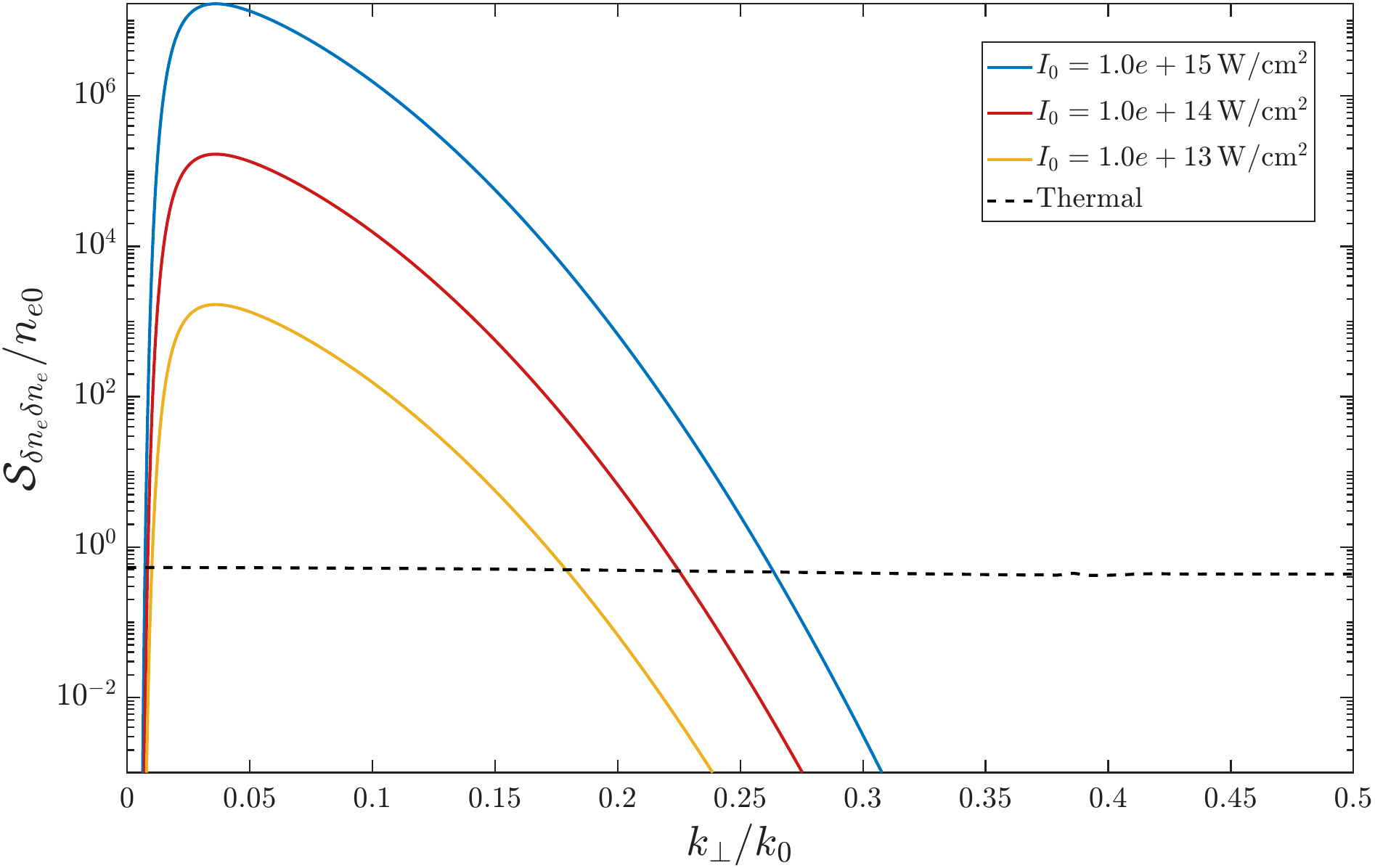}
\hfill
\includegraphics[width=0.49\linewidth]{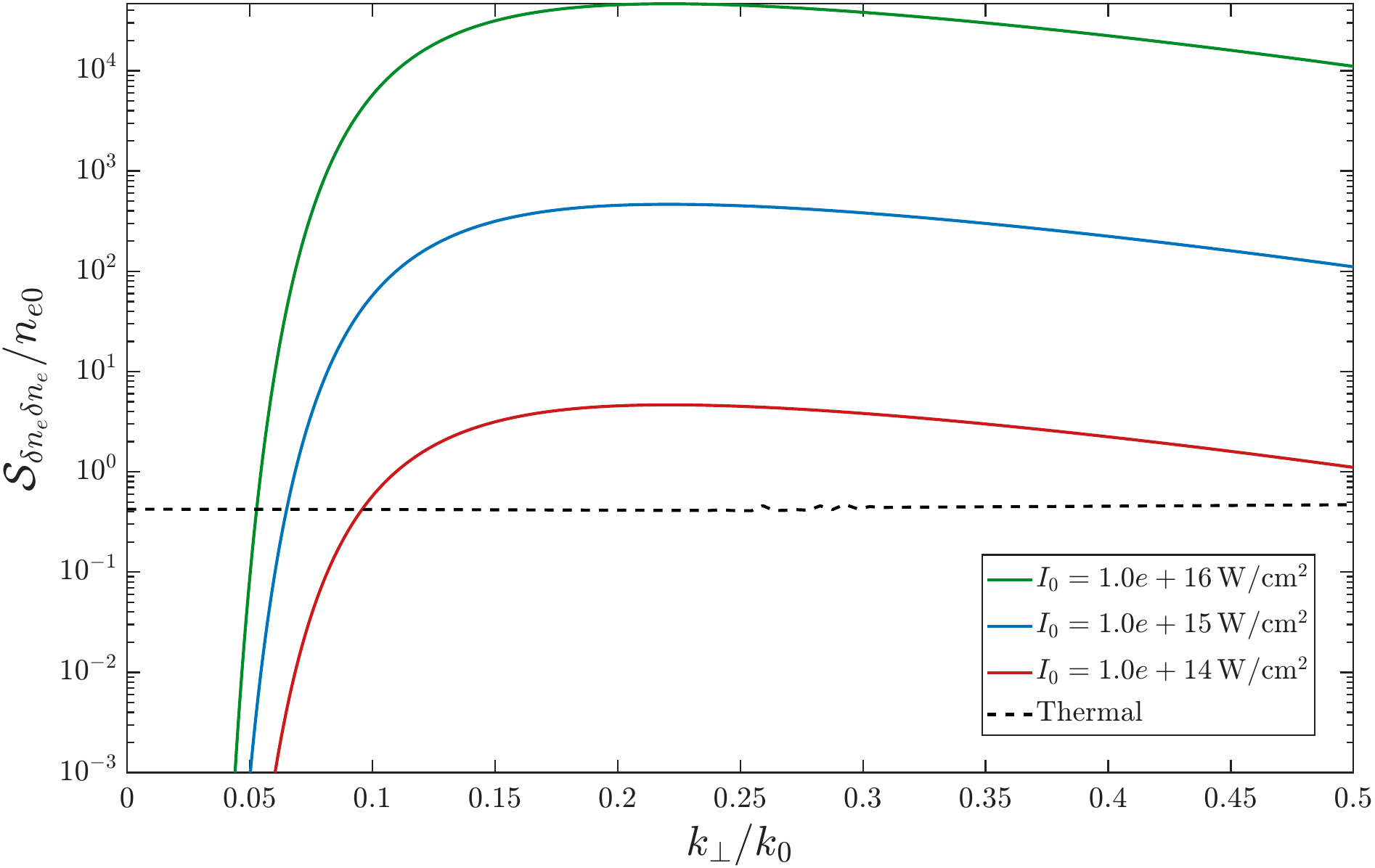}
\caption{We show a comparison between the speckle-driven fluctuations and thermal noise spectral densities through the unitless quantity $\mathcal{S}_{\delta n_e \delta n_e}/n_{e0}$ for different average intensities. In the left panel, we use the parameters $T_e=260\,\mathrm{eV}$, $T_i=130\,\mathrm{eV}$, $n_{e0}=2.00\times 10^{19}\,\mathrm{cm}^{-3}$, $Z=5$, $A=14$, $\lambda_0=351\,\mathrm{nm}$, and $L_{\perp}=1.72\,\mu\mathrm{m}$ mimicking a circular aperture RPP with "f-number" $\tilde{f}/\tilde{D}=6.7$. In the right panel, we use the parameters $T_e=2600\,\mathrm{eV}$, $T_i=1300\,\mathrm{eV}$, $n_{e0}=1.00\times 10^{20}\,\mathrm{cm}^{-3}$, $Z=2$, $A=4$, $\lambda_0=351\,\mathrm{nm}$, and $L_{\perp}=0.28\,\mu\mathrm{m}$ mimicking an annular RPP beam with an effective outer and inner diameter $\tilde{D}_O=6.2\,\mathrm{m}$, $\tilde{D}_I=2.6\,\mathrm{m}$, and focal length $\tilde{f}=6.0\,\mathrm{m}$. The left panel uses parameters representative of the Laboratory for Laser Energetics (LLE), while the right panel uses an effective annular parameterization motivated from the National Ignition Facility (NIF) beam configuration.}
\label{fig:spectral_density_compare}
\end{figure*}
As a comparison, the spectral density of density perturbations due to thermal noise, assuming Maxwellian distribution, in the same Fourier coordinates, can be written as \cite{froula2010plasma}
\begin{eqnarray}
S_{\delta n_e\delta n_e}^{\mathrm{NOISE}}\left(\mathbf{k},\omega\right) &=& \frac{2\pi n_{e0}}{k}\left(\left|\frac{1+\chi_i}{\varepsilon}\right|^2\sqrt{\frac{m_e}{2\pi T_e}}e^{-\frac{m_e}{2T_e}\frac{\omega^2}{k^2}}\right.\nonumber \\
 & &\left. +Z\left|\frac{\chi_e}{\varepsilon}\right|^2\sqrt{\frac{m_i}{2\pi T_i}}e^{-\frac{m_i}{2T_i}\frac{\omega^2}{k^2}}\right),
\label{thermal_spectral_density_density}
\end{eqnarray}
where $Z$ is the ion charge number defined by $q_i=-Zq_e$~\footnote{The form factor defined by Eq.~\eqref{thermal_spectral_density_density} differs from the standard definition of the $S(\mathbf{k},\omega)$ used in the book \cite{froula2010plasma} by the factor $n_{e0}$.}. To directly compare with the spectral density due to speckle density fluctuations from a RPP beam, we consider the integrated quantity
\begin{equation}
\mathcal{S}_{\delta n_e\delta n_e}^{\mathrm{NOISE}}(\mathbf{k}) := \frac{1}{2\pi}\int_{\mathbb{R}} S_{\delta n_e\delta n_e}^{\mathrm{NOISE}}(\mathbf{k},\omega)\,d\omega.
\end{equation}
We evaluate the spectra at $k_z=\ell_z^{-1}$, where $\ell_z=k_0L_\perp^2$ is the characteristic longitudinal speckle length. This selects the natural longitudinal scale of the RPP speckle pattern, allowing the thermal and speckle-driven density fluctuations to be compared at the same speckle-scale wavenumber. Figure~\ref{fig:spectral_density_compare} shows that, for experimentally relevant laser-plasma parameters, the speckle-driven contribution can exceed the thermal equilibrium fluctuation background by many orders of magnitude. This increase follows directly from the $I_0^2$ dependence of $\mathcal{S}_{\delta n_e\delta n_e}^{\mathrm{RPP}}$. Thus, in the stable sub-threshold regime, the relevant background density fluctuations need not be set by thermal particle noise alone; they may instead be dominated by stochastic density fluctuations driven by the speckle field.

\subsection{\label{sec:speckle_size_and_duration}Speckle Size and Annular Apertures}
The characteristic transverse and longitudinal speckle sizes are determined from the second-order spectral moments, as shown in Eq.~\eqref{eq:correlation_length}. The corresponding values for each aperture are listed in Table~\ref{table_speckle_sizes}. Such characteristic speckle dimensions are widely used to describe laser speckles and have proven useful in a variety of contexts, including effective speckle-volume estimates and reflectivity models~\cite{michel2023introduction,huller2010order}. The relationship between the longitudinal and transverse speckle sizes depends on the aperture shape. This is why the Gaussian width $L_{\perp}$ in Table~\ref{table_gaussian_beam_width} depends on whether the matching is performed in the transverse plane, a longitudinal slice, or the full three-dimensional volume: $L_{\perp}$ must preserve the Gaussian relation between longitudinal and transverse scales while matching the speckle volume of the other apertures. For the square, circular, and Gaussian apertures, once the transverse speckle size $\ell_\perp$ is fixed, the longitudinal speckle size $\ell_z$ is uniquely determined. The annular aperture, however, differs because varying the inner and outer spectral radii allows the transverse and longitudinal speckle sizes to be tuned more independently. This additional flexibility has also been exploited in optics to generate extended-depth-of-focus focal fields using annular apertures~\cite{hilden2023extended}. For example, taking $k_I\lesssim k_O$ narrows the annulus and increases $\ell_z$ relative to $\ell_\perp$, producing more elongated speckles. Taking $k_O\gg k_I$ instead produces shorter speckles, but still allows the longitudinal scale to be controlled through the choice of annular bandwidth. This provides greater flexibility than a circular aperture, where increasing the transverse bandwidth necessarily fixes both $\ell_\perp$ and $\ell_z$ through the $f$-number $(\tilde{f}/\tilde{D})$.

\begin{table}
\renewcommand{\arraystretch}{2}
\begin{ruledtabular}
\begin{tabular}{cccc}
 & $\ell_{\perp}$ & $\ell_z$ & $\ell_z(\ell_{\perp})$ \\
\hline
Square $\square$ 
& $\frac{\sqrt{3}}{k_c}$ 
& $\frac{3\sqrt{10}k_0}{2k_c^2}$ 
& $\frac{\sqrt{10}}{2}k_0\, \ell_\perp^2$ \\
Circular $\circ$
& $\frac{2}{k_c}$ 
& $\frac{4\sqrt{3}k_0}{k_c^2}$ 
& $\sqrt{3}k_0\, \ell_\perp^2$ \\
Annular $\circledcirc$
& $\frac{2}{\sqrt{k_O^2+k_I^2}}$ 
& $\frac{4\sqrt{3}k_0}{k_O^2-k_I^2}$ 
& $\sqrt{3}k_0 \frac{k_O^2+k_I^2}{k_O^2-k_I^2} \, \ell_{\perp}^{2}$ \\
Gaussian $g$
& $L_{\perp}$ 
& $k_0 L_{\perp}^2$ 
& $k_0\, \ell_{\perp}^2$ \\
\end{tabular}
\end{ruledtabular}
\caption{Characteristic transverse and longitudinal speckle sizes obtained from the second-order spectral moments for square, circular, annular, and Gaussian aperture spectra. The final column expresses the longitudinal speckle size in terms of the transverse speckle size.}
\label{table_speckle_sizes}
\end{table}

\section{Conclusion}
In this work, an asymptotic formula for the expected number of speckles above a prescribed intensity threshold was developed directly from the $\chi_2^2$ statistics of the intensity field. Beginning with the generalized beamlet representation of a phase-modulated beam in Eq.~\eqref{paraxial_solution}, a continuum random-field description of the laser envelope was constructed through its covariance function and spectral density, given in Eq.~\eqref{eq:covariance_spectral}. The central limit theorem then justifies treating $\mathcal{A}$ as a complex Gaussian random field, so that the intensity is $\chi_2^2$-distributed. Applying Kac-Rice theory directly to the $\chi_2^2$ intensity field yields the central result of this work: the maxima counting formula in Eq.~\eqref{maxima_count_chi_2}. The associated covariance functions, spectral densities, and second-order spectral moments were then evaluated for square, circular, annular, and Gaussian apertures. Comparisons with Monte Carlo simulations showed good agreement with the theoretical predictions and improved accuracy relative to the single-component ansatz used in previous approaches. These statistical results were then used to obtain several quantities of direct relevance to laser-plasma applications. In particular, the maxima counting formula was used to derive a simple average SBS reflectivity model, Eq.~\eqref{eq:reflectivity_formulation}, by estimating the contribution from intense speckles not captured by linear SBS estimates. The covariance calculations also provided a means of obtaining the spectral density of the speckle induced plasma density response, where a Gaussian envelope was used to obtain the closed-form approximation in Eq.~\eqref{eq:density_fluctuations_frequency_integrated}. Equivalent Gaussian beam widths were then tabulated in Table~\ref{table_gaussian_beam_width}, allowing this approximation to be matched to the statistics of the other aperture cases. The second-order spectral moments were used to define characteristic transverse and longitudinal speckle sizes, summarized in Table~\ref{table_speckle_sizes}, providing a compact reference across aperture geometries. The annular aperture also provided an additional degree of freedom for tailoring the characteristic speckle geometry. These results provide both a new maxima counting theory and a collection of practical statistical quantities that may be used directly when modelling speckle-driven laser-plasma processes.

\begin{acknowledgments}
We gratefully acknowledge D. Froula, S. H\"{u}ller, and A. Milder for helpful discussions and to K.R. McMillen for providing parameters used in Fig.~\ref{fig:spectral_density_compare} (left panel). Ian Min-Roberts and Wojciech Rozmus would like to acknowledge support from the U.S. Department of Energy (National Nuclear Security Administration) under Award No. DE-NA0004144: University of Rochester “National Inertial Confinement Fusion Program”. Work performed under the auspices of the U.S. Department of Energy by the Lawrence Livermore National Laboratory (LLNL) under Contract No. DE-AC52-07NA27344.
\end{acknowledgments}

\appendix
\section{Some Math Definitions}\label{sec:Some_Math_Definitions}
\textbf{Definition} ($\mathbb{R}^N$ Fourier Transform). We define the Fourier transform as $\mathcal{F}: L^2(\mathbb{R}^N) \to L^2(\mathbb{R}^N)$. Let $\mathbf{x}\in\mathbb{R}^N$ and $\mathbf{k}\in\mathbb{R}^N$, then using the non-unitary measure convention we define the Fourier transform for $f(\mathbf{x})\in L^2(\mathbb{R}^N)$ as
\begin{equation}\label{fourier_transform}
\hat{f}(\mathbf{k}) := \mathcal{F}[f](\mathbf{k}) = \int_{\mathbb{R}^N} f(\mathbf{x})e^{-i\mathbf{k}\cdot\mathbf{x}} \, d^N\mathbf{x}.
\end{equation}
Furthermore, there exists an inverse Fourier transform $\mathcal{F}^{-1}: L^2(\mathbb{R}^N) \to L^2(\mathbb{R}^N)$, for which $\hat{f}(\mathbf{k})\in L^2(\mathbb{R}^N)$ can be expressed as
\begin{equation}
f(\mathbf{x}) = \mathcal{F}^{-1}[\hat{f}](\mathbf{x}) = \frac{1}{(2\pi)^{N}}\int_{\mathbb{R}^N} \hat{f}(\mathbf{k}) e^{i\mathbf{k}\cdot\mathbf{x}} \, d^N\mathbf{k}.
\end{equation}

\textbf{Definition} (Random Field). Let $f\colon \mathbb{R}^N\to \mathbb{C}$ be a complex random field defined through the complex random measure $W$ by
\begin{equation}\label{complex_noise}
f(\mathbf{x}) = \int_{\mathbb{R}^N} e^{i\mathbf{k}\cdot\mathbf{x}}\, W(d\mathbf{k}),
\end{equation}
and assume $f$ is mean-zero.\\

\textbf{Definition} (Covariance). Let $f\colon \mathbb{R}^N\to \mathbb{C}$ be a mean-zero complex random field as defined in Eq.~\eqref{complex_noise}. Then for $\mathbf{x},\mathbf{x}^{\prime}\in\mathbb{R}^N$, the covariance $C_{ff}\colon\mathbb{R}^N\to\mathbb{C}$ is defined as
\begin{equation}\label{covariance_definition}
C_{ff}(\mathbf{x},\mathbf{x}^{\prime}) := \left\langle f(\mathbf{x}) \overline{f(\mathbf{x}^{\prime})} \right\rangle.
\end{equation}
If $f$ is homogeneous (or stationary), then the covariance depends only on the separation so we can shift our coordinates to one variable $\mathbf{x}-\mathbf{x}^\prime \mapsto \mathbf{x}$. Therefore, we write
\begin{equation}
C_{ff}(\mathbf{x}) = \left\langle f(\mathbf{x}) \overline{f(0)} \right\rangle.
\end{equation}

\textbf{Definition} (Spectral Density). Let $f\colon\mathbb{R}^N\to\mathbb{C}$ be a mean-zero, homogeneous complex random field as defined in Eq.~\eqref{complex_noise}. Let $\hat f:=\mathcal{F}[f]$. The spectral density $S_{ff}\colon \mathbb{R}^N\to\mathbb{R}_{\geq 0}$ is defined by
\begin{equation}\label{spectral_density_definition}
\left\langle \hat f(\mathbf{k})\overline{\hat f(\mathbf{k}')}\right\rangle := (2\pi)^N S_{ff}(\mathbf{k})\delta(\mathbf{k}-\mathbf{k}'),
\end{equation}
for $\mathbf{k},\mathbf{k}^{\prime}\in\mathbb{R}^N$. Since $f\colon\mathbb{R}^N\to\mathbb{C}$ is homogeneous with covariance $C_{ff}(\mathbf{x})$, then by the spectral representation theorem \cite[Theorem 5.4.2]{adler2007random}, the covariance and spectral density form a Fourier transform pair,
\begin{equation}
S_{ff}(\mathbf{k}) = \mathcal{F}[C_{ff}](\mathbf{k}).
\end{equation}
Equivalently, the covariance may be recovered through the inverse Fourier transform
\begin{equation}\label{covariance_inverse_fourier}
C_{ff}(\mathbf{x}) = \mathcal{F}^{-1}[S_{ff}](\mathbf{x}).
\end{equation}

\textbf{Definition} (Spectral Moments). Let $f\colon\mathbb{R}^N\to\mathbb{C}$ be a mean-zero, unit variance, homogeneous complex random field with spectral density $S_{ff}$. The $j$-th spectral moment tensor is defined by
\begin{equation}\label{spectral_moment_tensor_definition}
\Lambda^{(j)} := \frac{1}{(2\pi)^N} \int_{\mathbb{R}^N} \mathbf{k}^{\otimes j} S_{ff}(\mathbf{k})\,d^N\mathbf{k},
\end{equation}
where $\mathbf{k}^{\otimes j}$ denotes the $j$-fold tensor product of $\mathbf{k}$ with itself, so for example $\mathbf{k}^{\otimes 2} = \mathbf{k}\otimes \mathbf{k}$. Differentiating Eq.~\eqref{covariance_inverse_fourier} then evaluating at $\mathbf{x}=0$ yields
\begin{equation}
\Lambda^{(j)} = i^{-j}\nabla^{\otimes j} C_{ff}(0).
\end{equation}
For example the second-order spectral moments can be written as (in index notation),
\begin{equation}\label{second_order_spectral_moments_eq}
\Lambda_{ij}^{(2)} = - \left. \frac{\partial^2 C_{ff}(\mathbf{x})}{\partial x_i\partial x_j} \right\vert_{\mathbf{x}=0}.
\end{equation}
Assuming $C_{ff}$ is sufficiently smooth near $\mathbf{x}=0$, the covariance admits the local Taylor expansion
\begin{eqnarray}
C_{ff}(\mathbf{x}) &=& \sum_{j=0}^{\infty}\frac{1}{j!}\nabla^{\otimes j}C_{ff}(0)\left[\mathbf{x}^{\otimes j}\right]\nonumber\\
&=& \sum_{j=0}^{\infty}\frac{i^j}{j!}\Lambda^{(j)}\left[\mathbf{x}^{\otimes j}\right]
\label{cov_taylor_expansion}
\end{eqnarray}
where $\Lambda^{(j)}\left[\mathbf{x}^{\otimes j}\right]$ denotes the natural tensor contraction between the rank-$j$ tensor $\Lambda^{(j)}$ and the rank-$j$ tensor $\mathbf{x}^{\otimes j}$.

\section{Proof of Phase Corrected second-order Spectral Moment}\label{appendix:Proof_of_Effective_Second_Order_Spectral_Moment}
Let $C(\mathbf{x})$ be the unit-normalized covariance of $\mathcal{A}$, let $\mathbf{x}\in\mathbb{R}^N$, and expand $C(\mathbf{x})$ about $\mathbf{x}=0$ using Eq.~\eqref{cov_taylor_expansion},
\begin{align*}
C(\mathbf{x}) &= \sum_{j=0}^{\infty}\frac{i^j}{j!}\Lambda^{(j)}\left[\mathbf{x}^{\otimes j}\right]\\ &= 1 + i\, \Lambda^{(1)}\cdot \mathbf{x} - \frac{1}{2}\, \mathbf{x}^{T}\Lambda^{(2)}\mathbf{x} + \mathcal{O}\left(|\mathbf{x}|^3\right).
\end{align*}
Since a nonzero first spectral moment produces nonvanishing field gradient correlations, we define the phase-shifted field
\[
\mathring{\mathcal{A}}(\mathbf{x}) = \mathcal{A}(\mathbf{x}) e^{-i\, \Lambda^{(1)}\cdot \mathbf{x}},
\]
which preserves the intensity since $|\mathring{\mathcal{A}}(\mathbf{x})|^2 = |\mathcal{A}(\mathbf{x})|^2$, but yields a covariance with vanishing first order spectral moments as required for the conditional expectations in Worsley’s Kac-Rice formula \cite[Theorem 2.1]{worsley1994local}. So the new covariance is
\[
\mathring{C}(\mathbf{x}) = C(\mathbf{x})e^{-i\, \Lambda^{(1)}\cdot \mathbf{x}}.
\]
Expanding the original covariance and the exponent, gives us
\begin{eqnarray*}
\mathring{C}(\mathbf{x}) &=& \left[1 + i\, \Lambda^{(1)}\cdot \mathbf{x} - \frac{1}{2}\, \mathbf{x}^{T}\Lambda^{(2)}\mathbf{x} + \mathcal{O}\left(|\mathbf{x}|^3\right)\right]\\
&&\times \left[1 - i\, \Lambda^{(1)}\cdot \mathbf{x} - \frac{1}{2} \left(\Lambda^{(1)}\cdot \mathbf{x}\right)^2 + \mathcal{O}\left(|\mathbf{x}|^3\right)\right],
\end{eqnarray*}
then multiplying gives
\begin{eqnarray*}
\mathring{C}(\mathbf{x}) &=& 1 + \frac{1}{2}\left(\Lambda^{(1)}\cdot \mathbf{x}\right)^2 - \frac{1}{2}\, \mathbf{x}^{T}\Lambda^{(2)}\mathbf{x} + \mathcal{O}\left(|\mathbf{x}|^3\right)\\
&=& 1 - \frac{1}{2}\, \mathbf{x}^{T}\left(\Lambda^{(2)} - \Lambda^{(1)}\otimes\Lambda^{(1)}\right)\mathbf{x} + \mathcal{O}\left(|\mathbf{x}|^3\right).
\end{eqnarray*}
Therefore, the phase corrected second-order spectral moment $\mathring{\Lambda}^{(2)}$ is given by Eq.~\eqref{eq:effective_normalized_second_order_spectral_moment}, and $\mathring{\Lambda}^{(1)}=0$.

\section{Single Component Ansatz Counting Formula}\label{appendix:Garniers_Counting_Formula}
Garnier's~\cite{garnier1999statistics} asymptotic formula for the expected number of local maxima above a given intensity level set $u$, is listed as follows, in our convention~\footnote{Garnier multiplied his Gaussian Kac--Rice formula \cite[Eq.~(1)]{garnier1999statistics} by a factor of two because, under his convention, the covariance is halved by working with a single real Gaussian component of the underlying complex field.}:\\
For $\Omega \subset \mathbb{R}^2$:
\begin{align*}
\langle M_\Omega^{+}(u)\rangle =& \frac{2}{\pi^2}\,\lambda(\Omega)\,\det\left(\mathring{\Lambda}^{(2)}\right)^{1/2}\\
& \times \left(\left(\frac{1}{2}+\frac{\pi}{4}\right)u+\frac{1}{2}\right)e^{-u},
\end{align*}
for $\Omega \subset \mathbb{R}^3$:
\begin{align*}
\langle M_\Omega^{+}(u)\rangle =& \frac{10}{3\pi^{5/2}}\,\lambda(\Omega)\,\det\left(\mathring{\Lambda}^{(2)}\right)^{1/2}\\
& \times \left(u^{3/2}-\frac{3}{10}u^{1/2}\right)e^{-u},
\end{align*}
for $\Omega \subset \mathbb{R}^4$:
\begin{align*}
\langle M_\Omega^{+}(u)\rangle =& \frac{4}{\pi^3}\,\lambda(\Omega)\,\det\left(\mathring{\Lambda}^{(2)}\right)^{1/2}\\
& \times \left[\left(\frac{3\pi}{16}+\frac12\right)u^2-\left(\frac{3\pi}{8}+\frac12\right)u\right]e^{-u}.
\end{align*}

\section{Derivation of Kinetic Density Response Due to Speckles}\label{Derivation_of_Kinetic_Density_Response_Due_to_Speckles}
Starting from Vlasov's equation, we decompose the distribution function for each species $s\in S=\{e,i\}$ as $f_s(\mathbf{x},\mathbf{v},t) = f_{0s} + \delta f_{s}(\mathbf{x},\mathbf{v},t)$ where $f_{0s}$ is the equilibrium distribution and $\delta f_s$ is a small perturbation. The corresponding linearized kinetic equation is then
\[
\left(\frac{\partial}{\partial t}+\mathbf{v}\cdot\nabla\right)\delta f_s - \frac{q_s}{m_s}\nabla\delta\phi\cdot\nabla_{\mathbf{v}}f_{0s} - \frac{1}{m_s}\nabla U_s\cdot\nabla_{\mathbf{v}}f_{0s} = 0,
\]
where $U_s$ is the external ponderomotive drive potential, and $\delta \phi(\mathbf{x},t)$ is the electrostatic potential. The perturbative distribution $\delta f_s$ can be explicitly written in Fourier space as
\[
\delta \hat{f}_s(\mathbf{k},\mathbf{v},\omega) = -\frac{\mathbf{k} \cdot\nabla_{\mathbf{v}}f_{0s}(\mathbf{v})}{m_s\left(\omega-\mathbf{v}\cdot\mathbf{k}\right)}\left( q_s\delta\hat{\phi}(\mathbf{k},\omega) + \hat{U}_s(\mathbf{k},\omega)\right),
\]
where $\delta \hat{f}_s = \mathcal{F}[\delta f_s]$ for the Fourier transform $\mathcal{F}: L^2(\mathbb{R}^4)\to L^2(\mathbb{R}^4)$ in variables $(\mathbf{x},t)\mapsto (\mathbf{k},\omega)$. Hats on other transformed quantities are understood in the same sense. Defining density perturbations in Fourier space as 
\[
\delta \hat{n}_s(\mathbf{k},\omega) := \int_{\mathbb{R}^3} \delta \hat{f}_s(\mathbf{k},\mathbf{v},\omega) \, d^3\mathbf{v},
\]
and susceptibility as
\[
\chi_s(\mathbf{k},\omega) := \frac{q_s^2}{\varepsilon_0 m_s \lvert \mathbf{k} \rvert^2} \int_{\mathbb{R}^3}\frac{\mathbf{k} \cdot\nabla_{\mathbf{v}}f_{0s}(\mathbf{v})}{\omega-\mathbf{v}\cdot\mathbf{k}}\, d^3\mathbf{v},
\]
we get that
\begin{eqnarray*}
\delta \hat{n}_s(\mathbf{k},\omega) &=& -\frac{\varepsilon_0 \lvert \mathbf{k} \rvert^2}{q_s}\chi_s(\mathbf{k},\omega)\delta\hat{\phi}(\mathbf{k},\omega) \\
&& - \frac{\varepsilon_0 \lvert \mathbf{k} \rvert^2}{q_s^2}\chi_s(\mathbf{k},\omega)\hat{U}_s(\mathbf{k},\omega).
\end{eqnarray*}
Using Poisson's equation in Fourier space
\[
\lvert \mathbf{k} \rvert^2 \delta \hat{\phi} = \frac{1}{\varepsilon_0} \sum_{\alpha\in S} q_\alpha\delta \hat{n}_\alpha,
\]
we obtain the result
\begin{eqnarray*}
\delta \hat{n}_s(\mathbf{k},\omega) &=& \frac{\varepsilon_0 \lvert \mathbf{k} \rvert^2}{q_s}\left(\frac{1}{\varepsilon (\mathbf{k},\omega)}\sum_{\alpha\in S}\frac{\chi_\alpha(\mathbf{k},\omega)}{q_\alpha}\hat{U}_\alpha(\mathbf{k},\omega) \right. \\
&& \left.- \frac{1}{q_s}\hat{U}_s(\mathbf{k},\omega)\right)\chi_s(\mathbf{k},\omega),
\end{eqnarray*}
where $\varepsilon (\mathbf{k},\omega) := 1+\sum\limits_{\alpha\in S}\chi_\alpha(\mathbf{k},\omega)$ is the dielectric function. Since we are concerned with electron density response, we let $s=e$, and formulate our external ponderomotive drive potential as
\[
U_{e}(\mathbf{x},t) = \frac{q_e^2}{4m_e}\frac{2}{\varepsilon_0 c\, \omega_0^2} I,
\]
so that
\[
\delta \hat{n}_e(\mathbf{k},\omega) = -\frac{n_{e0}\lvert \mathbf{k} \rvert^2\lambda_{De}^2 }{2 c T_e n_c} \frac{1+\chi_i(\mathbf{k},\omega)}{\varepsilon(\mathbf{k},\omega)} \chi_e(\mathbf{k},\omega)\, \hat{I}(\mathbf{k},\omega),
\]
where $n_{e0}$ is the equilibrium electron density, $T_e$ is the electron temperature in energy units, $n_c=\varepsilon_0m_e\omega_0^2/q_e^2$ is the critical density, and $\lambda_{De}^2=\varepsilon_0T_e/(n_{e0}q_e^2)$ is the electron Debye length squared. Defining the response function as
\[
\hat{\mathcal{H}}(\mathbf{k},\omega) := \lvert \mathbf{k} \rvert^2\lambda_{De}^2 \frac{1+\chi_i(\mathbf{k},\omega)}{\varepsilon(\mathbf{k},\omega)} \chi_e(\mathbf{k},\omega).
\]
then evaluating the field at a second wavevector frequency pair $(\mathbf{k}^{\prime},\omega^{\prime})$, multiplying the two expressions together, and taking the ensemble average yields
\begin{eqnarray*}
\left\langle \delta \hat{n}_e(\mathbf{k},\omega) \overline{\delta \hat{n}_e(\mathbf{k}^{\prime},\omega^{\prime})} \right\rangle &=& \frac{n_{e0}^2}{4c^2T_e^2n_c^2} \hat{\mathcal{H}}(\mathbf{k},\omega) \overline{\hat{\mathcal{H}}(\mathbf{k}^{\prime},\omega^{\prime})} \\
&& \left\langle \hat{I}(\mathbf{k},\omega) \overline{\hat{I}(\mathbf{k}^{\prime},\omega^{\prime})} \right\rangle.
\end{eqnarray*}
Using the definition of spectral density in Eq.~\eqref{spectral_density_definition}, and integrating over the delta functions in the second wavevector frequency pair $(\mathbf{k}',\omega')$, gives Eq.~\eqref{eq:density_fluctuations}.

\bibliography{references}

\end{document}